\documentclass[prx,%
 reprint,
superscriptaddress,
 amsmath,amssymb,
 aps,
floatfix,
]{revtex4-1}
\usepackage{gensymb}
\usepackage{enumerate}
\usepackage{ulem}
\usepackage{orcidlink}
\usepackage{hyperref}
\hypersetup{colorlinks=true,linkcolor=blue,citecolor=blue,urlcolor=blue}
\usepackage{float}
\usepackage{graphicx}
\usepackage{dcolumn}
\usepackage{bm}

\usepackage{xcolor}

\usepackage{wrapfig}
\usepackage[left]{lineno}

\begin{document}

\preprint{APS/123-QED}

\title{Coherence-driven origin of metamagnetism in anisotropic heavy-fermion systems}

\author{Ewan Scott\,\orcidlink{0009-0002-8452-2262}}
\affiliation{Department of Mathematics, University College London, Gordon St., London WC1H 0AY, United Kingdom}

\author{Zheyu Wu\,\orcidlink{0009-0009-3839-5611}}
\affiliation{Cavendish Laboratory, University of Cambridge, JJ Thomson Av., Cambridge, CB3 0US}
\author{Theodore I. Weinberger\,\orcidlink{0000-0002-7769-7593}}
\affiliation{Cavendish Laboratory, University of Cambridge, JJ Thomson Av., Cambridge, CB3 0US}

\author{Alexander G. Eaton\,\orcidlink{0000-0001-6669-0807}}
\affiliation{Cavendish Laboratory, University of Cambridge, JJ Thomson Av., Cambridge, CB3 0US}

\author{Michal P. Kwasigroch\,\orcidlink{0000-0002-6613-2183}}
\affiliation{Department of Mathematics, University College London, Gordon St., London WC1H 0AY, United Kingdom}
\affiliation{Trinity College, Cambridge, CB2 1TQ, United Kingdom}
\date{\today}

\begin{abstract}

A number of heavy-fermion materials exhibit magnetic field-induced metamagnetism: on applying a field along the magnetic hard axis, the magnetization first rises gradually, then jumps abruptly once a critical field is reached. Despite decades of phenomenological modeling, the microscopic origin of the pronounced magnetic anisotropy underlying this behavior has remained unresolved. The same is true of a related, long-standing puzzle: an anomalous maximum in the hard-axis susceptibility versus temperature. Both are complicated in $5f$ compounds by the dual localized-itinerant character of the relevant electrons. Here we develop an analytic $c$--$f$ theory of magnetic anisotropy in heavy-fermion metamagnets, identifying the mixed susceptibility $\chi_{\rm cf}(T,h,p)$ as a single thermodynamic observable that unifies the anisotropic response across temperature, field, and pressure. We test this theory against primary and literature data for the heavy-fermion superconductor UTe$_2$, finding excellent quantitative agreement in the temperature, field, and pressure evolution of its magnetic anisotropy -- including a Kondo-coherence origin for the anomalous hard-axis susceptibility maximum, which we show is directly connected to the metamagnetic transition itself. Our results establish a general, microscopic, coherence-driven framework for anisotropic metamagnetism, applicable across the broad class of heavy-fermion compounds that display this phenomenology.

\end{abstract}

\maketitle

\section{Introduction}

Heavy-fermion compounds serve as an ideal playground for studying emergent quantum phases driven by strong electronic correlations~\cite{Coleman2015book}. In such systems, a delicate balance between localized $f$-electron moments and itinerant conduction electrons often yields highly anisotropic magnetic profiles. One particularly intriguing manifestation of such anisotropy is the frequent appearance of a pronounced maximum in the temperature-dependent magnetic susceptibility, for magnetic field $h$ aligned along the hard magnetic axis. Fundamentally, thermodynamic Maxwell relations posit that such a change in sign of the gradient of $\chi(T)$ signals an acute sensitivity of the entropy of the heavy Fermi liquid to $h$. Indeed, when $h$ is increased sufficiently along this hard direction, something remarkable occurs: the magnetization $M$ abruptly jumps upon crossing a metamagnetic phase boundary at $h_m$ (typically first-order in character), which can be accompanied by a dramatic reorganization of the magnetic anisotropy.

A number of heavy-fermion metamagnets follow this phenomenology - including UPt$_2$Si$_2$ \cite{BLECKMANN20102447,AMITSUKA1992173},  EuNi$_2$Pt$_2$ \cite{doi:10.7566/JPSJ.82.083708}, and superconducting UTe$_2$ \cite{ doi:10.7566/JPSJ.88.063706}.
In the lattermost case, the metamagnetism is particularly intriguing, as two magnetic field-induced superconducting states nestle against its phase boundary -- one stretching down to field strengths below $h_m$~\cite{Aoki_UTe2review2022,lewin2023review,Ranfieldboostednatphys2019,Knebel2019,Aoki_Hard,tony2024enhanced,zhang2025dimensionality,wu2026electricallycontrollablesuperconductingmemoryeffect}, and the other above~\cite{lewin2023review,Ranfieldboostednatphys2019,knafo2021comparison,helm2024,LANL_bulk_UTe2,frank2024orphan,tony2025brief,qcl,lewin2025halo,weinberger2025strangemetallicityencompasseshigh,wu2026directobservationspilloverhigh}. Understanding the microscopic character of the magnetic fluctuations present around $h_m$ -- and how they may or may not be responsible for pair-formation underpinning such exotic superconducting condensates -- are pertinent open questions~\cite{tokunaga2023longitudinal,TokiwaPRB2024,ripples,weinberger2026metamagnetismute2rolesitinerancy}.

Various approaches have been used to model heavy-fermion metamagnetism.  Since the 1960s, the seminal work of Wohlfarth and Rhodes \cite{Wohlfarth_Rhodes} has provided the foundation for understanding metamagnetism whenever itinerant electrons are present. Only itinerant moments are included in their analysis, and the metamagnetism, as well as the the maximum in the temperature-dependence of the susceptibility, are a consequence of the strong positive curvature of the density of states. A Ginzburg-Landau expansion in powers of the band magnetization can be motivated with a negative quartic coefficient that is responsible for the metamagnetism. The framework was later developed by Shimizu \cite{Shimizu_1982}, with spin-fluctuations incorporated by Yamada \cite{Yamada_1993}, who showed that whenever the metamagnetic jump is present at low temperature, there is a maximum in the temperature-dependence of $\chi(T)$.

Later models of metamagnetism incorporated localized as well as itinerant moments. They provide a more complete description of heavy-fermion metamagnetism, where in the crossover regime at intermediate temperatures the local moments are partially compensated and coexist with the heavy-fermion liquid, or where not all of the $f$-moments delocalize virtually even as $T\rightarrow0$. These include the duality model introduced by Miyake and Kuramoto \cite{MIYAKE_1991}, the work of Ono for a mixed-valent system \cite{Ono_1998}, the mean-field study of Ref. \cite{Kusminskiy_2008}, or the more recent studies in the context of UTe$_2$ \cite{Lacroix_5f2}. While they differ in the precise microscopic origin of the metamagnetic transition (MMT), either identifying it with the collapse of the entire heavy-fermion state \cite{Lacroix_5f2}, or the disappearance of the majority Fermi surface \cite{Kusminskiy_2008}, or the resurgence of the incoherent part of the $f$-electron \cite{Ono_1998}, except for the duality model of Ref. \cite{MIYAKE_1991}, they all involve Fermi surface reconstruction.  This heavy-fermion specific aspect is missing from the purely itinerant models of metamagnetism that follow the work of Wolfarth and Rhodes.

While present models capture many of metamagnetism's features, such as whenever there is a metamagnetic jump, there is also a hard-axis maximum at a temperature $T_{\rm max}$ that scales with $h_m$ \cite{Aoki_2013}, the microscopic origin of the huge anisotropy of the magnetic response has not been investigated in full.  In the original theory of Wolfarth and Rhodes, and in its later improvements, anisotropy is of secondary importance and the models were first applied to high-symmetry systems where the MMT occurs for all field directions, e.g., in Co based compounds \cite{Goto_1994}. This contrasts with heavy-fermion systems, where spin-orbit coupling and anisotropy are strong and metamagnetism is almost always confined to a narrow range of applied field directions \cite{Aoki_2013}. 

In this work, we bring anisotropy to the forefront of the observed metamagnetism and hard-axis maximum. We demonstrate that the interaction between localized and delocalized $f$-moments is particularly important in this respect.  While innocuous at high temperatures, we show that once coherence begins to develop, this interaction generates a sudden and rapid enhancement of magnetic anisotropy. We also take a closer look at the connection between the low-field and high-field responses and show that $T_{\rm max}$ and $h_m$ do not always track each other. This is easily demonstrated in UTe$_2$ by rotating the field away from the hard-axis towards one of the easy axes. While $T_m$ collapses, $h_m$ is strongly enhanced, and there is a range of angles where the metamagnetic jump is present but there is no susceptibility maximum. Nonetheless, the low and high field responses remain connected and, within a single microscopic model, we tie the metamagnetism of UTe$_2$ to its low-field behavior. In particular, we are able to accurately predict the enhancement of the metamagnetic field as it is rotated from the $b$-axis towards the $c$-axis solely from the low-field fits.

\begin{figure*}[t]
\includegraphics[width=\textwidth]{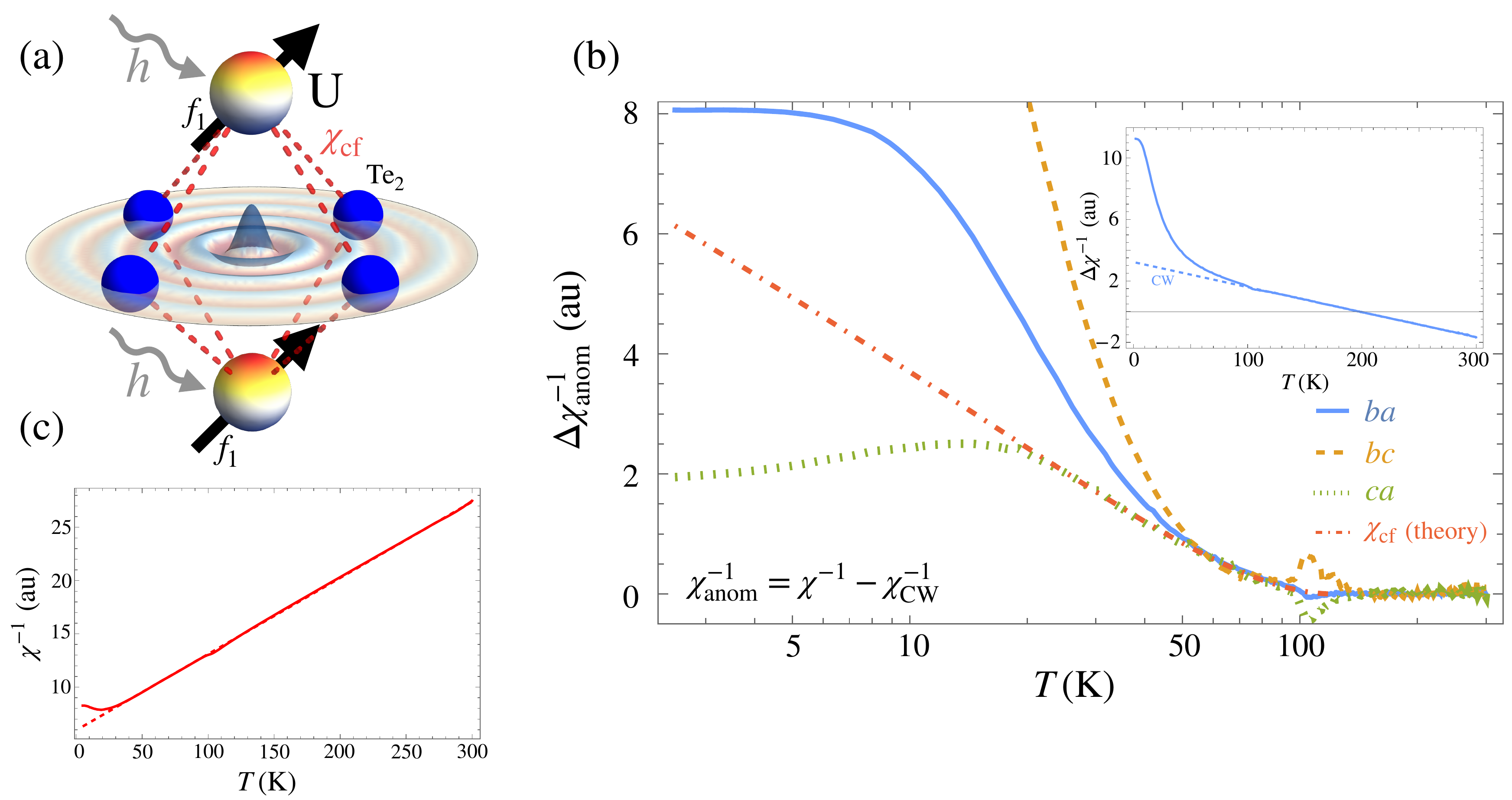}
    \centering
     \caption{
     (a) The mixed susceptibility, $\chi_{\rm cf}$, gives the  magnetization of the conduction electrons when a field is applied only to the hybridizing $f_1$-moments on the uranium ions. (b) The temperature evolution of the anomalous part of the inverse susceptibility $\chi^{-1}_{\rm anom}=\chi^{-1}-\chi^{-1}_{\rm CW}$ at ambient pressure.  The departure from the linear Curie-Weiss law $\chi^{-1}_{\rm CW}$ is most clearly seen when the difference $\Delta\chi^{-1}_{\rm anom}$ between two directions is taken. The legend gives the directions for which the difference is taken, $ba$ means $1/\chi^{b}_{\rm anom}-1/\chi^{a}_{\rm anom}$, etc. $\chi_{\rm cf}(T,h=0)\propto(1-T/T^*_{\rm coh})\ln(T^*_{\rm coh}/T)$  gives the empirically expected behavior for the mixed susceptibility with  $T^*_{\rm coh}$ fitted to 122~K. $\Delta\chi^{-1}_{\rm anom}$ was rescaled to collapse the data onto a single curve (details in the main text).   The inset shows the difference in the inverse susceptibilities for the $b$ and $a$ directions, $1/\chi^b-1/\chi^a$, with the dashed line showing the linear Curie-Weiss fit at high temperatures. There is a sudden deviation from Curie-Weiss behavior around $110{\rm K}$. 
     (c) Inverse susceptibility averaged over the three crystallographic directions $\chi^{-1}$ shows little deviation from Curie-Weiss behavior, unlike its anisotropy.}
     \label{fig:intro}
\end{figure*}

In Sec. \ref{sec:mix}, we introduce the single thermodynamic observable, the mixed susceptibility, that captures the coherence-driven evolution of the magnetic response, in particular its anisotropy, as a function of pressure, temperature and field. In Sec. \ref{sec:model}, we introduce the microscopic model that accounts for this coherence-driven evolution. In Sec. \ref{sec:low}, we characterize the low-field response, and test it against primary and secondary experimental data, whereas in Sec. \ref{sec:high}, we analyze the evolution of the response with field. We model the first-order MMT, and test our theory by predicting the evolution of $h_m$ in the $bc$-plane, solely from a low-field fit.

\section{Mixed susceptibility} \label{sec:mix}
Previous studies \cite{ Schmallian, Curro_QM,Curro_review} have identified the mixed susceptibility $\chi_{\rm cf}$ as a signature of the heavy Fermi liquid that develops below the coherence temperature. $\chi_{\rm cf}$ gives the susceptibility of the local $f$-moments to a magnetic field that  is applied only to conduction electrons. Because of the different hyperfine couplings with which nuclei couple to conduction electrons and local moments, the mixed susceptibility can be extracted through nuclear magnetic resonance (NMR) experiments \cite{Schmallian}. When an external magnetic field is applied, the internal field of the sample causes the nuclear precession frequencies to shift, in what is known as the Knight shift. Above the coherence temperature, when local $f$-moments and conduction electrons are only weakly coupled, this Knight shift has a linear dependence on the total susceptibility of the sample. However, below the coherence temperature, where conduction electrons and $f$-moments begin to hybridize strongly, the Knight shift deviates from the linear behavior, and this deviation is known as the anomalous Knight shift. It was shown in Ref. \cite{Schmallian} that the anomalous Knight shift is approximately proportional to the mixed susceptibility $\chi_{\rm cf}$ and the following logarithmic dependence below the coherence temperature $T^*_{\rm coh}$
\begin{align}
    \chi_{\rm cf}(T,h=0)\propto\left(1-\frac{T}{T^*_{\rm coh}}\right) \log\frac{T^*_{\rm coh}}{T}, \label{eq:log}
\end{align}
gave an excellent fit with experimental data across 14 heavy electron and mixed valent systems. Neglecting the small susceptibility of conduction electrons to a field that is only applied to them, $\chi_{\rm cc}$, the total uniform susceptibility of the heavy-fermion system can be written as the sum of the contribution from the heavy Fermi liquid ($2\chi_{\rm cf}$) and the Curie-Weiss contribution from the local $f$-moments ($\chi_{\rm ff}$): $\chi=2\chi_{\rm cf}+\chi_{\rm ff}$ \cite{Schmallian}. The mixed susceptibility directly measures the contribution from the heavy Fermi liquid component, whereas $\chi_{\rm ff}$ measures the local-moment contribution. As $f$-moments virtually hybridize below the coherence temperature, the former grows and the latter's effective Curie-Weiss moment falls in line with the respectively growing and decaying heavy Fermi liquid and local-moment components.

In this paper, we expand on the experimental signatures of the mixed susceptibility, beyond the anomalous Knight shift, and implicate it in the evolution of the magnetic anisotropy in UTe$_2$, as a function of temperature $T$, pressure $p$, and magnetic field $h$. We show that it is the dependence of the mixed susceptibility $\chi_{\rm cf}(T,h,p)$ on these thermodynamic variables that unifies the evolution of the magnetic anisotropy of UTe$_2$ across the $(T,h,p)$ phase space. We find that the mixed susceptibility is not only proportional to the deviation of the Knight shift from linear dependence, but also the deviation of the anisotropy of the inverse susceptibility from the linear Curie-Weiss law, which we refer to as the anomalous part to emphasize the close analogy.

\section{Microscopic model}\label{sec:model}

We are motivated by the 2-fluid model of Nakatsuji, Pines and Fisk (NPF) \cite{NPF_model}, where much of the phenomenology of heavy-fermions can be explained by including contributions from local $f$-moments as well as the coherent heavy Fermi liquid made up of hybridized $f$-moments and conduction electrons. Further, the dual nature of $5f$ electrons in uranium heavy-fermion systems has been highlighted in various experimental studies, such as X-ray spectroscopy that observed local-moment signatures even in highly itinerant Pauli paramagnets \cite{Amorese_2020}, or inelastic neutron scattering \cite{Lee_2018}. The above studies underscore the importance of including multiplet states besides band physics to fully characterize uranium intermetallics.   We therefore include both hybridizing and non-hybridizing $f$-electrons in our microscopic description of UTe$_2$, with the latter not participating in the Fermi liquid and retaining a local moment down to the lowest temperatures.  

\begin{figure*}
\includegraphics[width=\textwidth]{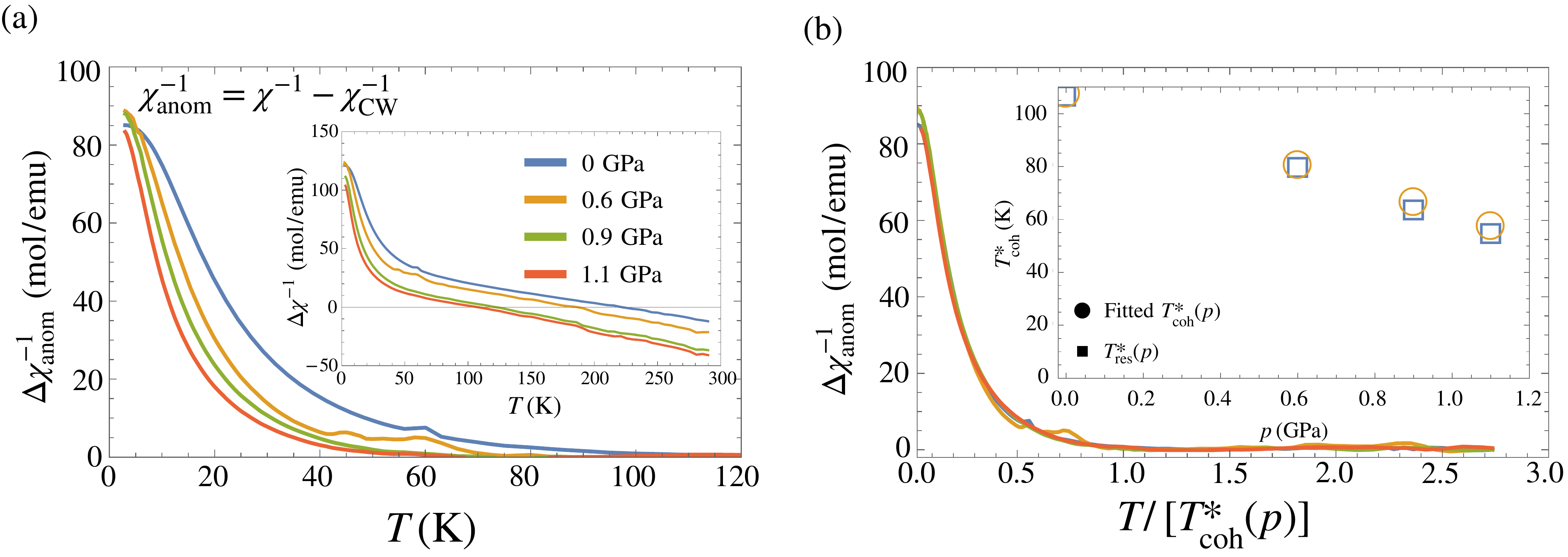}
    \centering
     \caption{The temperature evolution of the anomalous part of the susceptibility $\chi^{-1}_{\rm anom}=\chi^{-1}-\chi^{-1}_{\rm CW}$ for pressure in the range $0-1.1 \;{\rm GPa}$. The anomalous evolution is most evident if the difference between two directions is taken. The difference between $a$ and $b$ directions $\Delta_{}\chi^{-1}_{\rm anom}=1/\chi^{b}_{\rm anom}-1/\chi^{a}_{\rm anom}$ is taken (similar behaviour observed for other directions). (a) $\Delta_{}\chi^{-1}_{\rm anom}$ is plotted as a function of temperature. The inset shows the difference in the inverse susceptibility along the $b$ and $a$ axes $1/\chi^{b}_{}-1/\chi^{a}_{}$ before the subtraction of the high-temperature Curie-Weiss fit. (b) An appropriate coherence temperature $T^{\ast}_{\rm coh}(p)$ was chosen for each pressure to collapse the experimental data onto a single curve.  The inset shows the fitted $T^*_{\rm coh}(p)$ as well as the resistivity maximum temperature extracted from Ref. \cite{Thomas_2020} for each pressure. Susceptibility data from Ref. \cite{Akoi}, kindly provided by D. Aoki, was used. (Linear interpolation was used if the susceptibility was measured at slightly different temperatures for each direction.)}
     \label{fig:pressure}
\end{figure*}

UTe$_2$ is a multi $f$-electron system and is believed to be somewhere between $5f^2$ and $5f^3$ electronic configurations with pressure driving the system closer to $5f^2$ \cite{Amorese_2020}. A recent X-ray scattering study \cite{christovam2024stabilization} highlighted $5f^2$ as a good starting point for describing low-energy phenomena. We will focus on the case where each uranium contains two $f$-electrons ($f_1$ and $f_2$) that form stable spin-$1/2$ moments. We will assume that, because of the different occupied orbitals, the $f_1$-moment interacts strongly with conduction electrons, virtually hybridizes and participates in the Fermi surface, whereas the $f_2$-moment interacts weakly and remains localized. Similar models have been used previously to capture the duality of $5f$-electrons of UTe$_2$ \cite{Lacroix_5f2}. 

The Kondo interaction between the $f$-moments and conduction electrons can be written as
\begin{eqnarray}
     H_K = J_1\sum_{i}  \mathbf{S}_{1}(\mathbf{r}_i)\cdot \mathbf{S}_{c}(\mathbf{r}_i)+J_2 \sum_{i} \mathbf{S}_{2}(\mathbf{r}_i)\cdot \mathbf{S}_{c}(\mathbf{r}_i),
\end{eqnarray}
where $i$ indexes the lattice sites and $\mathbf{S}_c$ and $\mathbf{S}_{1,2}$ are the conduction and $f_{1,2}$-moment spins, respectively. We take $J_1\rho\sim 1$ and $J_2\rho\ll 1$ with $\rho$ the density of states at the Fermi level, so that the $f_1$-moment fractionalizes and effectively adds charge to the Fermi volume, whereas the $f_2$-moment remains a local spin-$1/2$ degree of freedom. This is in line with the Doniach argument where, because of competition with the RKKY interaction, the formation of the heavy-fermion state is only favored at sufficiently strong coupling. Another possible scenario is that of underscreening, where the two $f$-moments compete for the same conduction electron. In this case, it suffices for $J_1 \gtrsim J_2$ and $J_1\rho\sim 1$, for the $f_1$-moment to virtually hybridize and the $f_2$-moment to remain localized.

An {\it effective} anisotropy is endowed to the spin-$1/2$ degrees of freedom, through a combination of spin-orbit coupling and crystal electric fields (CEF), that breaks the degeneracy of the uranium $5f^2$-subspace. In the presence of orthorhombic symmetry, the most general anisotropy in our model is given by
\begin{eqnarray}
    H_{\rm anis}= \sum_{i} D^{\eta} S_{1}^{\eta}(\mathbf{r}_{i})S_{2 }^{\eta}(\mathbf{r}_{i}), \label{eq:perp}
\end{eqnarray}
where $D^{\eta}=\{D^a,D^b,D^c\}$ is the local orthorhombic anisotropy. Although we are focusing on anisotropy, the average of $D^{\eta}$ does not need to be zero.
The first order correction to the total energy of the system coming from the anisotropy is given by 
\begin{align}
    \langle H_{\rm anis}\rangle 
    &=\sum_{\eta,i} D^\eta 
    \bar{ S}_1(\mathbf{r}_i) 
    \bar{S}_2(\mathbf{r}_i) 
    \nonumber\\
    &=
   - \sum_{\eta,i,j} J_2D^\eta \bar{S}^\eta_{2}(\mathbf{r}_i)\chi_{\rm cf}(\mathbf{r}_i-\mathbf{r}_j)\bar{S}^\eta_2(\mathbf{r}_j), \label{eq:anis}
\end{align}
where we have treated the $f_2$-moments within the static (classical) approximation, and $\bar{S}^\eta_{1,2}(\mathbf{r}_i)$ is the expectation value of the $f_{1,2}$-spin on lattice site $i$. $\chi_{\rm cf}(\mathbf{r}_i-\mathbf{r}_j)$ is the zero-field susceptibility of the hybridized $f_1$-moment on site $i$, to a field that is applied to a conduction electron orbital on site $j$ only. We can see that the mixed susceptibility governs the anisotropy of the energy landscape of the unhybridized $f_2$-moments.
Making a mean-field approximation for the $f_2$-moments, we can find their susceptibility
\begin{align}
    \chi^{\eta}(T,h=0):&=\left.\frac{\partial \bar{S}^\eta_2(T,h)}{\partial h}\right|_{h=0}
     \nonumber\\&=\frac{1}{4T-\lambda-2J_2D^\eta\chi_{\rm cf}(T,h=0)}, \label{eq:CW}
\end{align}
where we have neglected the direct action of the field on the band, which is valid as long as $h\ll J_2 \bar{S}_2^{\eta}$, and $\lambda<0$ is the antiferromagnetic and isotropic molecular field felt by the $f_2$-moments, given by the RKKY interaction energy $\sum_{\mathbf{r}\neq \mathbf{0}}J_2^2\chi_{\rm cc}(\mathbf{r})$ at weak coupling. $\chi_{\rm cc}(\mathbf{r})$ is the zero-field, zero-temperature susceptibility of a conduction electron orbital to a field that is applied to a conduction electron orbital a displacement $\mathbf{r}$ away, and $\chi_{\rm cf} (T,h=0)=\sum_{i}\chi_{\rm cf}(\mathbf{r}_i)$. Motivated by the weak temperature dependence of $\chi_{\rm cc}$, which is in congruence with the anomalous Knight shift experiments across many heavy-fermion compounds \cite{Schmallian}, as well as calculations on the half-filled periodic Anderson model \cite{Curro_review}, we will neglect the temperature, and later also field, dependence of $\lambda$ and treat it as a constant.

\section{Low-field response} \label{sec:low}
\begin{figure*}[t]
    \centering
\includegraphics[width=\textwidth]{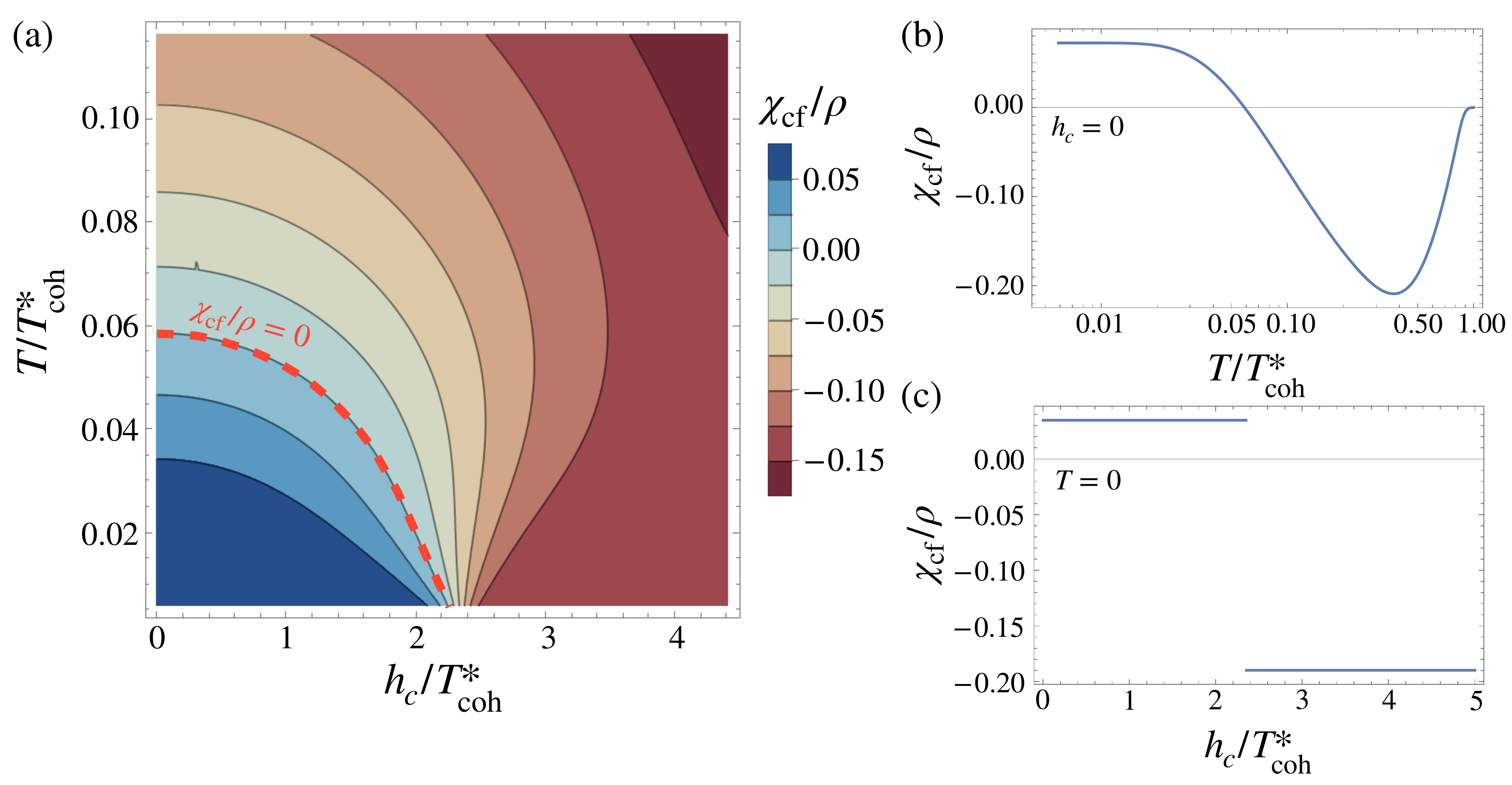}
     \caption{(a) Mixed susceptibility $\chi_{\rm cf}(T,h_c)$, as a function of temperature $T$ and field $h_c$ applied only to conduction electrons. The mixed susceptibility gives the susceptibility of the hybridized $f_1$-moment when the field is only applied to conduction electrons. The calculation is performed in the large-$N$ limit of the Kondo model with a constant density of states $\rho$ for the conduction band, and for a representative conduction electron filling $n_c=0.8$ and Kondo coupling $2J_1\rho=1$. The coherence temperature $T^*_{\rm coh}$ has been estimated as the temperature at which the hybridization field turns on. (b) The zero-field mixed susceptibility $\chi_{\rm cf}(T,h_c=0)$, as a function of temperature $T$. (c) The zero-temperature mixed susceptibility $\chi_{\rm cf}(T=0,h_c)$, as a function of field $h_c$ applied only to conduction electrons.}
     \label{fig:mixed_x}
\end{figure*}
Focusing on the local-moment contribution to the total susceptibility, Eq. \ref{eq:CW}, shows that the average inverse susceptibility approximately follows the Curie-Weiss law (neglecting the temperature dependence of the isotropic molecular field and assuming that the average $D^{\eta}$ is small), but the anisotropic part tracks the evolution of the mixed susceptibility and therefore begins to deviate strongly from Curie-Weiss law below the coherence temperature
\begin{align}
    \langle \chi^{-1}\rangle :&=\frac{1}{3}\sum_{\eta} 1/\chi^{\eta}=4T-\lambda,
    \nonumber\\
    \Delta_{\eta\eta'}\chi^{-1}:&=1/\chi^{\eta}-1/\chi^{\eta'}
    \nonumber\\
    &=-2J_2(D^{\eta}-D^{\eta'})\chi_{\rm cf}(T,h=0) \label{eq:anom}.
\end{align}
Note that we have not explicitly included orbital angular momentum in our analysis. CEF fields and spin-orbit coupling will naturally generate anisotropy in the molecular field experienced by, as well as effective moments carried by, the total angular momentum of the uranium ions.  These anisotropies will not give rise to the observed departure from Curie-Weiss behavior. Further, a generic CEF scheme can indeed be responsible for non-Curie-Weiss behavior at temperatures comparable to the CEF level splittings as thermal depopulation of CEF levels takes place, but is unlikely to give a sudden enhancement of the anisotropy that begins precisely at the coherence temperature.  We argue that this enhancement is instead caused by the coherence-amplified interaction between the spin degrees of freedom of hybridized and unhybridized $f$-moments. 

Eq. \ref{eq:anom} shows excellent agreement with the low-field response of  UTe$_2$, presented in Fig. \ref{fig:intro}, where the direction-averaged inverse susceptibility follows the Curie-Weiss law to very low temperatures but the anisotropic part deviates strongly at significantly higher temperatures. The anomalous deviation of the inverse susceptibility from the linear Curie-Weiss law will be denoted by $\chi^{-1}_{\rm anom}:=\chi^{-1}-\chi^{-1}_{\rm CW}$.  Fig. \ref{fig:intro}(b) looks at the difference in the anomalous part for two directions $\eta,\eta'$: $\Delta_{\eta\eta'}\chi^{-1}_{\rm anom}:=  1/\chi^{\eta}_{\rm anom}-1/\chi^{\eta'}_{\rm anom}$. This quantity rapidly turns on around the same temperature for all $\eta,\eta'$ of $T^*_{\rm coh}=122$~K, which is close to the measured resistivity maximum along the $a$-axis, extrapolated to zero pressure \cite{Thomas_2020}. In accordance with Eq. \ref{eq:anom}, $\Delta_{\eta\eta'}\chi^{-1}_{\rm anom}$ was divided by a scaling factor proportional to $D^{\eta}-D^{\eta'}$ to collapse the three curves onto $\chi_{\rm cf}(T,h=0)$. The scaling factors used give the following constraint: $D^a=2.92D^c-1.92D^b$.  We note that $\Delta_{ca}\chi^{-1}_{\rm anom}$ follows the logarithmic behavior expected from the mixed susceptibility for much longer than $\Delta_{ba,bc}\chi^{-1}_{\rm anom}$.  We suspect that the quenching of the local-moment contribution to magnetization along the $b$-direction at low temperatures leads to an exponentially fast temperature evolution of $\Delta_{ba,bc}\chi^{-1}_{\rm anom}$. The quenching could be caused by the strong interaction between the closest uranium moments that form the dimers in UTe$_2$ if it has a high $b$ vs. $ac$ anisotropy.

Applying hydrostatic pressure reduces the coherence temperature, as UTe$_2$ is believed to move from $5f^3$ to $5f^2$ valency \cite{Amorese_2020}, leading to weaker hybridization of $f$-moments. Fig. \ref{fig:pressure} shows $\Delta_{ba}\chi^{-1}_{\rm anom}$ at a range of pressures. The anomalous part of the anisotropy of the inverse susceptibility can be collapsed onto a single curve by simply rescaling temperature (no vertical scaling has been applied). The inset of Fig. \ref{fig:pressure}(b) shows that the rescaling temperature $T^*_{\rm coh}$ follows closely the resistivity maximum temperature $T^*_{\rm res}$ over the pressure range of $0-1.1\;{\rm GPa}$, demonstrating that the rapid growth of magnetic anisotropy, away from Curie-Weiss behavior, is directly driven by coherence. 

We note that Eq. \ref{eq:CW} shows that the logarithmic development of the mixed susceptibility can generate a maximum in the susceptibility as a function of temperature along an axis with $D^{\eta}<0$ (assuming $\partial \chi_{\rm cf} (T,h=0)/\partial T>0$). Since a maximum is only observed for the $b$-direction in UTe$_2$, we take $D^b<0, D^{a,c}>0$.  The temperature at which it occurs, denoted by $T_{\rm max}$, satisfies $\partial \chi (T_{\rm max},h=0)/\partial T=0$ and using Eq. \ref{eq:log} scales as $T_{\rm max} \propto -J_2D^b$. We will see later that the field-dependence of the zero-temperature mixed susceptibility will give rise to a metamagnetic transition within our model, only if $D^{\eta}<0$, i.e. only for large fields applied along the $b$-direction, which is in agreement with experimental observation. 

\section{High field response}\label{sec:high}
 
 We now go beyond the expected behavior of the mixed susceptibility as a function of temperature and pressure (which tunes the coherence temperature), and include an external magnetic field. We have calculated the mixed susceptibility within the large-$N$ approximation \cite{NRead_1983}, where the Kondo exchange between the conduction electron and the $f_1$-moment is decoupled in the hybridization channel and the mean hybridisation field $V=\frac{1}{2}\sum_{\sigma}\langle c^{\dagger}_{\sigma}(\mathbf{r}_i )f_{1\sigma}(\mathbf{r}_i )\rangle$ is self-consistently determined, subject to a global constraint that there is one $f_1$-electron per lattice site. $f_{1\sigma}$ and $c_{\sigma}$ are, respectively, the usual fermionic annihilation operators for $f_1$ and conduction electrons with spin $\sigma$.  Fig. \ref{fig:mixed_x} shows our results for $\chi_{\rm cf}(T,h_c)$ for a representative conduction electron filling (number per lattice site) of $n_c=0.8$, and dimensionless Kondo coupling $2J_1\rho=1$. The density of states in the conduction band has been approximated by a constant $\rho$ and $h_c$ is an external magnetic field coupling to conduction electrons only (similar behavior is observed when the field also couples to the hybridized $f_1$-moments). We see that the mixed susceptibility not only has a rapid temperature-evolution below $T_{\rm coh}^*$ (estimated here as the temperature at which the hybridization field $V$ turns on), but that this evolution also involves a change of sign that is not captured by the empirically motivated Eq. \ref{eq:log}. We note that mean-field theory is not able to capture the logarithmic rise which is expected to be generated by corrections arising from the fluctuations of the hybridization field. The temperature-induced sign change of $\chi_{\rm cf}(T,h_c)$ signals an inversion of anisotropy at a scale that tracks the coherence temperature. Indeed, UTe$_2$ experiences a crossing of the uniform $a$-axis and $b$-axis susceptibilities at a temperature directly proportional to the coherence temperature over the pressure range $0-1.4 \;{\rm GPa}$ \cite{Scott_2025}. We also note that the zero-temperature mixed susceptibility experiences a jump and a sign change when the external magnetic field $h_c$ exceeds a critical value, signaling another reversal of anisotropy, and a return to the high-temperature easy-direction. In the case of UTe$_2$, this corresponds to the hard $b$-axis becoming the easy-axis and it is therefore tempting to explore the connection between this sudden reversal of the mixed susceptibility and the metamagnetic transition that is observed when a high field is applied to the UTe$_2$ hard-axis. We note here that the jump in the zero-temperature mixed susceptibility is linked to the spin-selective insulator transition, that has been observed in dynamical mean-field theory calculations \cite{Peters_2012}, when parts of the strongly magnetized band become insulating.
\begin{figure*}
\includegraphics[width=0.95\textwidth]{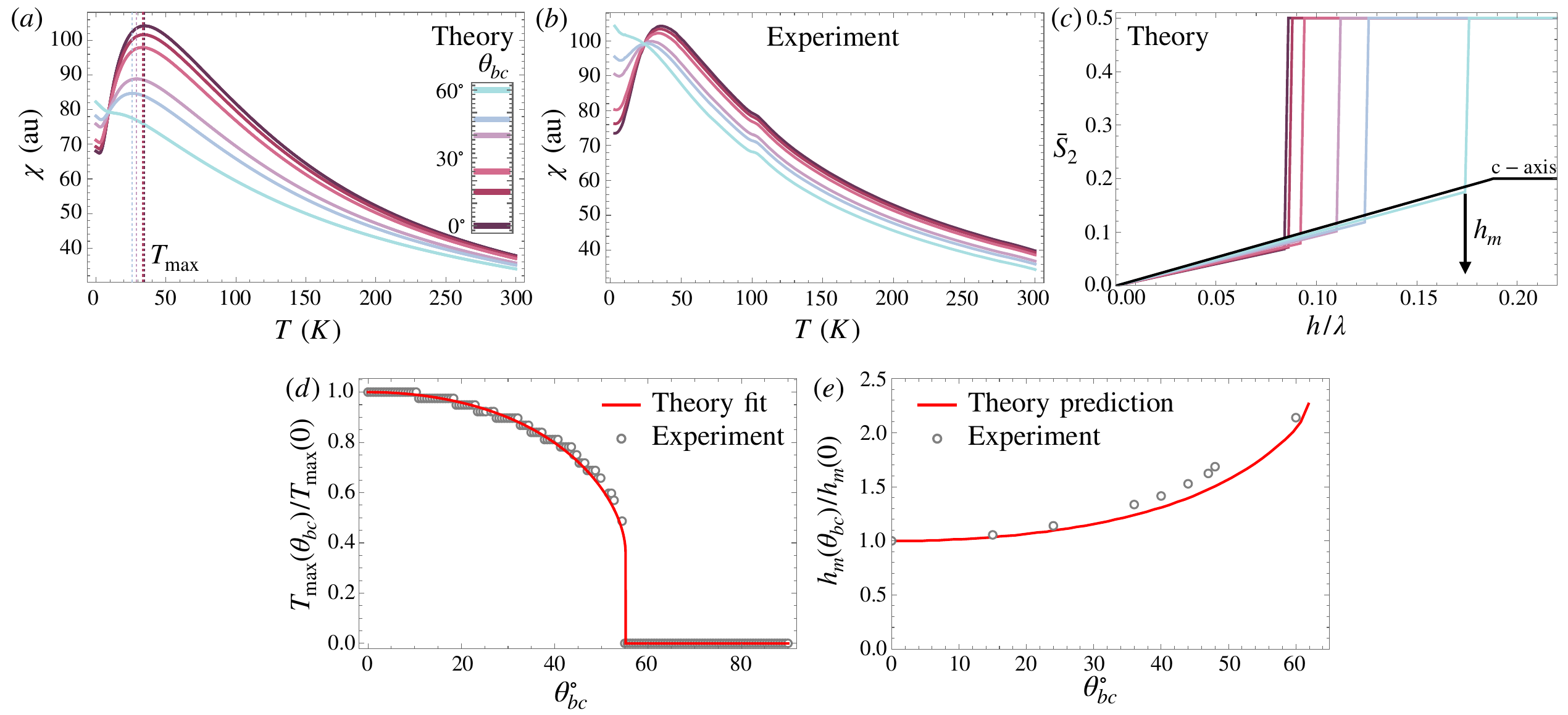}
    \centering
     \caption{(a) The theoretically calculated zero-field susceptibility $\chi(T,h=0)$ as a function of temperature for field applied at an angle $\theta_{bc}$ from the $b$-axis towards the $c$-axis in the range $0\degree$ to $60 \degree$.  The mean-field approximation in Eq. \ref{eq:CW} was used with the following representative parameters: Kondo couplings $2J_1\rho=1,2J_2\rho=0.99$, molecular field $\lambda\rho=-0.25$, conduction electron filling $n_c=0.9$. The anisotropy parameters of $D^b\rho=-0.85$, $D^{c}\rho=0.19$ were tuned to obtain the best possible fit for the susceptibility maximum temperature $T_{\rm max}(\theta_{\rm bc})$. (b) The experimentally obtained susceptibility for the same range of angles $\theta_{bc}$. (c) The theoretically calculated  zero-temperature magnetization of the unhybridized $f_2$-moment $\bar{S}_2$ as a function of field $h$ applied at an angle $\theta_{bc}$. The mean-field approximation in Eq. \ref{eq:MFmet} and the same parameters as above were used. (d) The theoretically calculated and experimentally measured $T_{\rm max}(\theta_{bc})$ as a function of $\theta_{bc}$. Grey circles are the experimentally measured values and the red line is the theoretical fit. (e) The theoretically calculated critical field $h_m(\theta_{\rm bc})$ at which the metamagnetic jump occurs as a function of $\theta_{bc}$.  Grey circles are the experimentally measured values~\cite{qcl,qcldata} and the line is the theoretical prediction based on the parameters obtained from the low-field fit presented in panels (a) and (d).}
\label{fig:meta}
\end{figure*}

\subsection{Metamagnetic transition}\label{sec:jump}

We model the metamagnetic transition by again neglecting the direct application of the external field $h$ to the heavy-fermion band and treating the unhybridized $f_2$-moments within the static (classical) approximation. The total magnetic energy of the $f_2$-moments as a function  of their uniform magnetization $\bar{\mathbf{S}}_2$ is given by
\begin{align}\nonumber
    E(\mathbf{\bar{S}}_2)=-\frac{1}{2}\lambda |\mathbf{\bar{S}}_2|^2-&h\begin{pmatrix}
        \cos(\theta_{b\gamma})\\
        \sin(\theta_{b\gamma})
    \end{pmatrix}\cdot\mathbf{\bar{S}}_2\,+\\&
    D^b\bar{S}_1^b\cdot \bar{S}_2^b+D^\gamma\bar{S}_1^\gamma\cdot \bar{S}_2^\gamma, \label{eq:MFmet}
\end{align}
where
\begin{equation}
    \mathbf{\bar{S}}_1= \left\{  \begin{matrix} 
     \chi_1 (-J_2 \mathbf{\bar{S}}_2)&|\mathbf{\bar{S}}_2|<S^*\\
        \chi_2 (-J_2\mathbf{\bar{S}}_2 )+\frac{\mathbf{\bar{S}}_2}{|\mathbf{\bar{S}}_2|}(\chi_2 -\chi_1)J_2 S^{*}&|\mathbf{\bar{S}}_2|\geq S^*
    \end{matrix}\right.,\label{eq:S}
\end{equation}
$\gamma\in\{a,c\}$ is one of the easy axes, and $\theta_{b\gamma}$ is the angle from the $b$-axis towards the $\gamma$-axis at which the field is applied ($\theta_{b\gamma}=0\degree$ corresponds to a field along the $b$ axis). At zero temperature, $\chi_{\rm cf}(T=0, h_c)$ is a step function with $\chi_{1/2}$ equal to $\chi_{\rm cf}(T=0, h_c)$ each side of the step that takes place when the effective magnetic field applied to the conduction electrons by the unhybridized $f_2$-moments via the Kondo coupling, $\mathbf{h}_c=-J_2\mathbf{\bar{S}}_2$, reaches a critical field value of $h^{\ast}_{c}=-J_2S^*$. That is, when the $f_2$-moments reach a critical polarization of $S^*$. The $\frac{\mathbf{\bar{S}}_2}{|\mathbf{\bar{S}}_2|}(\chi_2 -\chi_1)J_2 S^{*}$ term in Eq. \ref{eq:S} ensures that the magnetization of the heavy-fermion band, specifically the $f_1$ hybridized moment, is a continuous function of the local moment magnetization $\mathbf{\bar{S}}_2$. The equilibrium magnetization of the $f_2$-moment at zero-temperature is obtained by numerically minimizing Eq. \ref{eq:MFmet}. When the field is oriented along an axis with $D^{\eta}<0$, the $f_2$-moment magnetization experiences a jump across the critical polarization of $S^*$, at a critical field denoted by $|\mathbf{h}|=h_m$. The energy, which is bilinear in $\mathbf{\bar{S}}_2$, develops a minimum for $|\mathbf{\bar{S}}_2|>S^*$ which is lower than the minimum for $|\mathbf{\bar{S}}_2|<S^*$, because of a more ferromagnetic total molecular field. (For some parameters, the lowest value of the energy in the allowed range $|\mathbf{\bar{S}}_2|\leq\frac{1}{2}$ is at the full magnetization of the spin-1/2 moment.)  No such jump is favorable if the field is applied along an axis with $D^{\eta}>0$ (the total molecular field is more antiferromagnetic at $|\mathbf{\bar{S}}_2|>S^*$) and the system is stuck at the minimum at $S^*$ until much higher fields. There is no Fermi surface reconstruction, and consequently, no jump in the mixed susceptibility.

One may wonder how the energy landscape of the unhybridized $f_2$-moment given in Eq. \ref{eq:MFmet} relates to the previously postulated Ginzburg-Landau expansions of Shimizu \cite{Shimizu_1982} and Yamada \cite{Yamada_1993} where the fourth power of magnetization has a negative coefficient. Firstly, it is important to note the perturbative and mean-field limitations of the current theory. Eq. \ref{eq:MFmet} is only valid sufficiently far away from $|\mathbf{S}_2|=S^{\ast}$, where we can be sure beyond-mean-field fluctuations or the perturbative interaction in Eq. \ref{eq:perp} do not affect which side of the MMT and jump in the mixed susceptibility the system is on. In the current theoretical approximation the energy minima at $|\mathbf{S}_2|<S^{\ast}$ and $|\mathbf{S}_2|>S^{\ast}$ are thus disconnected and should be thought of as separate local approximations for $E(\mathbf{S}_2)$. There is a kink in the free energy at $|\mathbf{S}_2|=S^{\ast}$, which we expect fluctuations and higher order corrections in $D^{\eta}$ to smooth out. Consequently, the sharp  plateauing of $f_2$-magnetization that is observed if the field is applied along an axis with $D^{\eta}>0$ will also be smoothed out. Moving away from zero temperature will also have a significant effect. At any non-zero temperature the step at $h_c^{\ast}$ in $\chi_{\rm cf}(T,h)$ is smoothed out. For small $|\mathbf{S}_2|$, we can approximate $\chi_{\rm cf}(T,h_c)=\chi_{\rm cf}(T,h_c=0) + \frac{1}{2}\frac{\partial^2 \chi_{\rm cf}}{\partial h^2}(T,h_c=0)J_2^2|\mathbf{S}_2|^2 $ and the negative curvature $\frac{\partial^2 \chi_{\rm cf}}{\partial h^2}(T,h_c=0)<0$ at the level of a phenomenological Ginzburg-Landau theory performs the same role as the negative curvature of the band susceptibility in the duality model of Miyake and Kuramoto \cite{MIYAKE_1991}, where it is a consequence of the negative curvature of the density of states at the Fermi level $N''(E_F)<0$. It directly leads to an $|\mathbf{S}_2|^4$ term in the energy with a negative coefficient.

Fig. \ref{fig:meta} presents the low and high field response to a magnetic field applied at an angle $\theta_{bc}$ from the b-axis towards the c-axis. Experimental data is presented together with theoretical calculations for representative values of the Kondo coupling and molecular field $2J_1\rho=1,2J_2\rho=0.99,\lambda\rho=-0.25$ and filling $n_c=0.9$. The mean-field approximations in Eq. \ref{eq:CW} and Eq. \ref{eq:MFmet} were used and the mixed susceptibility was calculated using the large-$N$ limit of the Kondo model, with a constant density of states approximation. The anisotropy parameters of $D^b\rho=-0.85$ and $D^{c}\rho=0.19$ were tuned to obtain the best possible fit for the temperature of the susceptibility maximum $T_{\rm max}(\theta_{bc})$ as a function of $\theta_{bc}$. We emphasize here that we can obtain an excellent fit for a range of the isotropic parameters $J_1\rho,J_2\rho$ and $\lambda\rho$, by tuning just these two anisotropy parameters in each case. This again shows the central role of the anisotropy interaction between localized and itinerant $f$-moments, given in Eq. \ref{eq:anis}, in shaping the evolution of the susceptibility maximum.  Panels (a) and (b) respectively show the theoretical and experimentally obtained zero-field susceptibility as a function of temperature for angles $\theta_{bc}$ in the range $0{\degree}$ to $60{\degree}$. The experimentally measured susceptibility is obtained by superposing the susceptibility measured for fields applied along the principal axes. Panel (c) presents the theoretically calculated magnetization of the unhybridized $f_2$-moment as a function of applied field $h$ for angles in the same range. Panel (d) shows $T_{\rm max}(\theta_{bc})$, extracted from the experimental data in panel (b), as well as the theoretical fit. In panel (e), the experimentally measured critical magnetic field $h_m$ at which the metamagnetic jump occurs is plotted as a function of $\theta_{bc}$, together with the theoretical prediction, calculated with the above parameters, optimized to obtain the best fit for $T_{\rm max}(\theta_{bc})$. It is remarkable that we are able to accurately predict the evolution of $h_m(\theta_{bc})$ just using the low-field fit. The metamagnetic jump vanishes around $\theta_{bc}=65\degree$ in the theoretical calculation, although another one develops at higher fields after initial plateauing of $|\bar{\mathbf{S}}_2|$ at $|\bar{\mathbf{S}}_2|=S^{\ast}$, and eventually disappears around $\theta_{bc}=85\degree$.

One can obtain a good understanding of the results presented in Fig. \ref{fig:meta} by working to first order in the anisotropy $D^{\eta}$. In this approximation the magnetic field at which the metamagnetic transition takes place and the temperature of the susceptibility maximum are given by
\begin{align}
    h_m(\theta_{b\gamma})&=-\lambda S^{*}-\frac{1}{2}D(\theta_{b\gamma})J_2S^*(3\chi_1+\chi_2)
    \nonumber\\
    T_{\rm max}(\theta_{b\gamma}) &\propto J_2 D(\theta_{b\gamma})
\end{align}
for $D(\theta_{b\gamma})=D^b\cos^2\theta_{b\gamma}+D^{\gamma}\sin^2\theta_{b\gamma}<0$. From the large-$N$ calculation, we have found that $(3\chi_1+\chi_2)<0$ for not-too-strong Kondo coupling $J_1$. Consequently, the harder the direction (i.e. more negative $D(\theta_{b\gamma})$), the lower the metamagnetic field $h_m$. The maximum $h_m$, equal to -$\lambda S^{*}$, is attained at the critical angle $\theta_{b\gamma}=\arctan(\sqrt{-D^b/D^{\gamma}})$ when the effective anisotropy $D(\theta_{b\gamma})$, the metamagnetic jump and the susceptibility maximum temperature all vanish. We note that to first order in $D^{\eta}$ this maximum metamagnetic field is direction independent. Because the critical angle at which the metamagnetic jump vanishes is higher for a rotation in the $bc$-plane than for a rotation in the $ba$-plane, it means that $h_m$ has to rise faster with angle for the former, from its lowest value at $\theta_{b\gamma}=0$ to its highest value at the critical angle. The dependence of $h_m$ on $D(\theta_{b\gamma})$ should be contrasted with that of $T_{\rm max}$. While the former increases with angle (when $D(\theta_{b\gamma})$ becomes less negative), the latter decreases with angle, which indeed is the experimental observation and shows that the two do not need to track each other. Our model predicts both to fall with pressure, as they indeed do, since the Kondo coupling $J_{1,2}$ is expected to fall. 

We have not been able to obtain a good fit for $T_{\rm max}(\theta_{ba})$ or $h_m(\theta_{ba})$ for a field applied at angle $\theta_{ba}$ from the $b$-axis towards the $a$-axis. (The data can be found in Appendix \ref{app:ba}.) Presumably, this is because we have taken effective moments and molecular constants to be isotropic, and this is a much worse approximation for the $ba$-plane than for the $bc$-plane \cite{Akoi}. However, we can estimate $D^{a}\rho\approx 2.2$ using the constraint from Sec. \ref{sec:low} and the fitted values of $D^{b,c}$. Using the small anisotropy approximation, this gives critical angles of $\theta_{ba}\approx 32\degree$ and $\theta_{bc}\approx 65\degree$ for when the susceptibility maxima and metamagnetic jumps should vanish, which are not far from the experimentally measured values.

\section{Conclusion}

We have identified the interaction between the spin degrees of freedom of hybridized and unhybridized $f$-moments as the  source of the rapid and sudden enhancement of the magnetic anisotropy that begins at the coherence temperature. We have demonstrated that the mixed susceptibility $\chi_{\rm cf}(T,h,p)$ is a single thermodynamic observable that can characterize this coherence-driven evolution of the  magnetic anisotropy  across the $(T,h,p)$ phase space.  Its logarithmic growth below the coherence temperature is responsible for the sudden deviation of the magnetic susceptibility of UTe$_2$ from Curie-Weiss behavior, which is most apparent if the difference between inverse susceptibilities measured along two directions is taken. This has been verified with experimental data across the  $0-1.1\;{\rm GPa}$ pressure range, which allows us to tune the coherence temperature. The logarithmic growth is also responsible for the maximum in the temperature dependence of the zero-field susceptibility that is seen along the hard-direction. The step-change in the field dependence of $\chi_{\rm cf}$ at zero-temperature, on the other hand, is responsible for the metamagnetic jump along the hard-direction. We test our theory by looking at the evolution of the susceptibility maximum and the metamagnetic jump as the applied field is rotated from the $b$-axis towards the $c$-axis. We find that we can obtain an excellent fit for the temperature of the susceptibility maximum as a function of the rotation angle and accurately predict the evolution of the critical field at which the metamagnetic jump takes place as as function of the angle using parameters obtained from the low-field fit.

In summary, we have shown that magnetic anisotropy is at the heart of the magnetic response of the heavy-fermion superconductor UTe$_{2}$. This includes the maximum in the temperature dependence of the hard-axis magnetization at low fields, as well as a jump in the hard-axis magnetization at high fields. Our results will be applicable to many heavy-fermion compounds that exhibit similar phenomenology.

\bibliography{apssamp}

\begin{thebibliography}{48}%
\makeatletter
\providecommand \@ifxundefined [1]{%
 \@ifx{#1\undefined}
}%
\providecommand \@ifnum [1]{%
 \ifnum #1\expandafter \@firstoftwo
 \else \expandafter \@secondoftwo
 \fi
}%
\providecommand \@ifx [1]{%
 \ifx #1\expandafter \@firstoftwo
 \else \expandafter \@secondoftwo
 \fi
}%
\providecommand \natexlab [1]{#1}%
\providecommand \enquote  [1]{``#1''}%
\providecommand \bibnamefont  [1]{#1}%
\providecommand \bibfnamefont [1]{#1}%
\providecommand \citenamefont [1]{#1}%
\providecommand \href@noop [0]{\@secondoftwo}%
\providecommand \href [0]{\begingroup \@sanitize@url \@href}%
\providecommand \@href[1]{\@@startlink{#1}\@@href}%
\providecommand \@@href[1]{\endgroup#1\@@endlink}%
\providecommand \@sanitize@url [0]{\catcode `\\12\catcode `\$12\catcode `\&12\catcode `\#12\catcode `\^12\catcode `\_12\catcode `\%12\relax}%
\providecommand \@@startlink[1]{}%
\providecommand \@@endlink[0]{}%
\providecommand \url  [0]{\begingroup\@sanitize@url \@url }%
\providecommand \@url [1]{\endgroup\@href {#1}{\urlprefix }}%
\providecommand \urlprefix  [0]{URL }%
\providecommand \Eprint [0]{\href }%
\providecommand \doibase [0]{http://dx.doi.org/}%
\providecommand \selectlanguage [0]{\@gobble}%
\providecommand \bibinfo  [0]{\@secondoftwo}%
\providecommand \bibfield  [0]{\@secondoftwo}%
\providecommand \translation [1]{[#1]}%
\providecommand \BibitemOpen [0]{}%
\providecommand \bibitemStop [0]{}%
\providecommand \bibitemNoStop [0]{.\EOS\space}%
\providecommand \EOS [0]{\spacefactor3000\relax}%
\providecommand \BibitemShut  [1]{\csname bibitem#1\endcsname}%
\let\auto@bib@innerbib\@empty
\bibitem [{\citenamefont {Coleman}(2015)}]{Coleman2015book}%
  \BibitemOpen
  \bibfield  {author} {\bibinfo {author} {\bibfnamefont {P.}~\bibnamefont {Coleman}},\ }\href {https://doi.org/10.1017/CBO9781139020916} {\emph {\bibinfo {title} {Introduction to many-body physics}}}\ (\bibinfo  {publisher} {Cambridge University Press},\ \bibinfo {year} {2015})\BibitemShut {NoStop}%
\bibitem [{\citenamefont {Bleckmann}\ \emph {et~al.}(2010)\citenamefont {Bleckmann}, \citenamefont {Otop}, \citenamefont {Süllow}, \citenamefont {Feyerherm}, \citenamefont {Klenke}, \citenamefont {Loose}, \citenamefont {Hendrikx}, \citenamefont {Mydosh},\ and\ \citenamefont {Amitsuka}}]{BLECKMANN20102447}%
  \BibitemOpen
  \bibfield  {author} {\bibinfo {author} {\bibfnamefont {M.}~\bibnamefont {Bleckmann}}, \bibinfo {author} {\bibfnamefont {A.}~\bibnamefont {Otop}}, \bibinfo {author} {\bibfnamefont {S.}~\bibnamefont {Süllow}}, \bibinfo {author} {\bibfnamefont {R.}~\bibnamefont {Feyerherm}}, \bibinfo {author} {\bibfnamefont {J.}~\bibnamefont {Klenke}}, \bibinfo {author} {\bibfnamefont {A.}~\bibnamefont {Loose}}, \bibinfo {author} {\bibfnamefont {R.}~\bibnamefont {Hendrikx}}, \bibinfo {author} {\bibfnamefont {J.}~\bibnamefont {Mydosh}}, \ and\ \bibinfo {author} {\bibfnamefont {H.}~\bibnamefont {Amitsuka}},\ }\href {\doibase https://doi.org/10.1016/j.jmmm.2010.02.054} {\bibfield  {journal} {\bibinfo  {journal} {Journal of Magnetism and Magnetic Materials}\ }\textbf {\bibinfo {volume} {322}},\ \bibinfo {pages} {2447} (\bibinfo {year} {2010})}\BibitemShut {NoStop}%
\bibitem [{\citenamefont {Amitsuka}\ \emph {et~al.}(1992)\citenamefont {Amitsuka}, \citenamefont {Sakakibara}, \citenamefont {Sugiyama}, \citenamefont {Ikeda}, \citenamefont {Miyako}, \citenamefont {Date},\ and\ \citenamefont {Yamagishi}}]{AMITSUKA1992173}%
  \BibitemOpen
  \bibfield  {author} {\bibinfo {author} {\bibfnamefont {H.}~\bibnamefont {Amitsuka}}, \bibinfo {author} {\bibfnamefont {T.}~\bibnamefont {Sakakibara}}, \bibinfo {author} {\bibfnamefont {K.}~\bibnamefont {Sugiyama}}, \bibinfo {author} {\bibfnamefont {T.}~\bibnamefont {Ikeda}}, \bibinfo {author} {\bibfnamefont {Y.}~\bibnamefont {Miyako}}, \bibinfo {author} {\bibfnamefont {M.}~\bibnamefont {Date}}, \ and\ \bibinfo {author} {\bibfnamefont {A.}~\bibnamefont {Yamagishi}},\ }\href {\doibase https://doi.org/10.1016/0921-4526(92)90090-F} {\bibfield  {journal} {\bibinfo  {journal} {Physica B: Condensed Matter}\ }\textbf {\bibinfo {volume} {177}},\ \bibinfo {pages} {173} (\bibinfo {year} {1992})}\BibitemShut {NoStop}%
\bibitem [{\citenamefont {Hiranaka}\ \emph {et~al.}(2013)\citenamefont {Hiranaka}, \citenamefont {Nakamura}, \citenamefont {Hedo}, \citenamefont {Takeuchi}, \citenamefont {Mori}, \citenamefont {Hirose}, \citenamefont {Mitamura}, \citenamefont {Sugiyama}, \citenamefont {Hagiwara}, \citenamefont {Nakama},\ and\ \citenamefont {Ōnuki}}]{doi:10.7566/JPSJ.82.083708}%
  \BibitemOpen
  \bibfield  {author} {\bibinfo {author} {\bibfnamefont {Y.}~\bibnamefont {Hiranaka}}, \bibinfo {author} {\bibfnamefont {A.}~\bibnamefont {Nakamura}}, \bibinfo {author} {\bibfnamefont {M.}~\bibnamefont {Hedo}}, \bibinfo {author} {\bibfnamefont {T.}~\bibnamefont {Takeuchi}}, \bibinfo {author} {\bibfnamefont {A.}~\bibnamefont {Mori}}, \bibinfo {author} {\bibfnamefont {Y.}~\bibnamefont {Hirose}}, \bibinfo {author} {\bibfnamefont {K.}~\bibnamefont {Mitamura}}, \bibinfo {author} {\bibfnamefont {K.}~\bibnamefont {Sugiyama}}, \bibinfo {author} {\bibfnamefont {M.}~\bibnamefont {Hagiwara}}, \bibinfo {author} {\bibfnamefont {T.}~\bibnamefont {Nakama}}, \ and\ \bibinfo {author} {\bibfnamefont {Y.}~\bibnamefont {Ōnuki}},\ }\href@noop {} {\bibfield  {journal} {\bibinfo  {journal} {Journal of the Physical Society of Japan}\ }\textbf {\bibinfo {volume} {82}},\ \bibinfo {pages} {083708} (\bibinfo {year} {2013})}\BibitemShut {NoStop}%
\bibitem [{\citenamefont {Miyake}\ \emph {et~al.}(2019)\citenamefont {Miyake}, \citenamefont {Shimizu}, \citenamefont {Sato}, \citenamefont {Li}, \citenamefont {Nakamura}, \citenamefont {Homma}, \citenamefont {Honda}, \citenamefont {Flouquet}, \citenamefont {Tokunaga},\ and\ \citenamefont {Aoki}}]{doi:10.7566/JPSJ.88.063706}%
  \BibitemOpen
  \bibfield  {author} {\bibinfo {author} {\bibfnamefont {A.}~\bibnamefont {Miyake}}, \bibinfo {author} {\bibfnamefont {Y.}~\bibnamefont {Shimizu}}, \bibinfo {author} {\bibfnamefont {Y.~J.}\ \bibnamefont {Sato}}, \bibinfo {author} {\bibfnamefont {D.}~\bibnamefont {Li}}, \bibinfo {author} {\bibfnamefont {A.}~\bibnamefont {Nakamura}}, \bibinfo {author} {\bibfnamefont {Y.}~\bibnamefont {Homma}}, \bibinfo {author} {\bibfnamefont {F.}~\bibnamefont {Honda}}, \bibinfo {author} {\bibfnamefont {J.}~\bibnamefont {Flouquet}}, \bibinfo {author} {\bibfnamefont {M.}~\bibnamefont {Tokunaga}}, \ and\ \bibinfo {author} {\bibfnamefont {D.}~\bibnamefont {Aoki}},\ }\href {\doibase 10.7566/JPSJ.88.063706} {\bibfield  {journal} {\bibinfo  {journal} {Journal of the Physical Society of Japan}\ }\textbf {\bibinfo {volume} {88}},\ \bibinfo {pages} {063706} (\bibinfo {year} {2019})}\BibitemShut {NoStop}%
\bibitem [{\citenamefont {Aoki}\ \emph {et~al.}(2022)\citenamefont {Aoki}, \citenamefont {Brison}, \citenamefont {Flouquet}, \citenamefont {Ishida}, \citenamefont {Knebel}, \citenamefont {Tokunaga},\ and\ \citenamefont {Yanase}}]{Aoki_UTe2review2022}%
  \BibitemOpen
  \bibfield  {author} {\bibinfo {author} {\bibfnamefont {D.}~\bibnamefont {Aoki}}, \bibinfo {author} {\bibfnamefont {J.~P.}\ \bibnamefont {Brison}}, \bibinfo {author} {\bibfnamefont {J.}~\bibnamefont {Flouquet}}, \bibinfo {author} {\bibfnamefont {K.}~\bibnamefont {Ishida}}, \bibinfo {author} {\bibfnamefont {G.}~\bibnamefont {Knebel}}, \bibinfo {author} {\bibfnamefont {Y.}~\bibnamefont {Tokunaga}}, \ and\ \bibinfo {author} {\bibfnamefont {Y.}~\bibnamefont {Yanase}},\ }\href {\doibase 10.1088/1361-648X/AC5863} {\bibfield  {journal} {\bibinfo  {journal} {J. Phys. Condens. Matter}\ }\textbf {\bibinfo {volume} {34}},\ \bibinfo {pages} {243002} (\bibinfo {year} {2022})}\BibitemShut {NoStop}%
\bibitem [{\citenamefont {Lewin}\ \emph {et~al.}(2023)\citenamefont {Lewin}, \citenamefont {Frank}, \citenamefont {Ran}, \citenamefont {Paglione},\ and\ \citenamefont {Butch}}]{lewin2023review}%
  \BibitemOpen
  \bibfield  {author} {\bibinfo {author} {\bibfnamefont {S.~K.}\ \bibnamefont {Lewin}}, \bibinfo {author} {\bibfnamefont {C.~E.}\ \bibnamefont {Frank}}, \bibinfo {author} {\bibfnamefont {S.}~\bibnamefont {Ran}}, \bibinfo {author} {\bibfnamefont {J.}~\bibnamefont {Paglione}}, \ and\ \bibinfo {author} {\bibfnamefont {N.~P.}\ \bibnamefont {Butch}},\ }\href {https://doi.org/10.1088/1361-6633/acfb93} {\bibfield  {journal} {\bibinfo  {journal} {Rep. Prog. Phys.}\ } (\bibinfo {year} {2023})}\BibitemShut {NoStop}%
\bibitem [{\citenamefont {Ran}\ \emph {et~al.}(2019)\citenamefont {Ran}, \citenamefont {Liu}, \citenamefont {Eo}, \citenamefont {Campbell}, \citenamefont {Neves}, \citenamefont {Fuhrman}, \citenamefont {Saha}, \citenamefont {Eckberg}, \citenamefont {Kim}, \citenamefont {Graf}, \citenamefont {Balakirev}, \citenamefont {Singleton}, \citenamefont {Paglione},\ and\ \citenamefont {Butch}}]{Ranfieldboostednatphys2019}%
  \BibitemOpen
  \bibfield  {author} {\bibinfo {author} {\bibfnamefont {S.}~\bibnamefont {Ran}}, \bibinfo {author} {\bibfnamefont {I.~L.}\ \bibnamefont {Liu}}, \bibinfo {author} {\bibfnamefont {Y.~S.}\ \bibnamefont {Eo}}, \bibinfo {author} {\bibfnamefont {D.~J.}\ \bibnamefont {Campbell}}, \bibinfo {author} {\bibfnamefont {P.~M.}\ \bibnamefont {Neves}}, \bibinfo {author} {\bibfnamefont {W.~T.}\ \bibnamefont {Fuhrman}}, \bibinfo {author} {\bibfnamefont {S.~R.}\ \bibnamefont {Saha}}, \bibinfo {author} {\bibfnamefont {C.}~\bibnamefont {Eckberg}}, \bibinfo {author} {\bibfnamefont {H.}~\bibnamefont {Kim}}, \bibinfo {author} {\bibfnamefont {D.}~\bibnamefont {Graf}}, \bibinfo {author} {\bibfnamefont {F.}~\bibnamefont {Balakirev}}, \bibinfo {author} {\bibfnamefont {J.}~\bibnamefont {Singleton}}, \bibinfo {author} {\bibfnamefont {J.}~\bibnamefont {Paglione}}, \ and\ \bibinfo {author} {\bibfnamefont {N.~P.}\ \bibnamefont {Butch}},\ }\href {\doibase 10.1038/s41567-019-0670-x} {\bibfield  {journal} {\bibinfo  {journal} {Nat. Phys.}\
  }\textbf {\bibinfo {volume} {15}},\ \bibinfo {pages} {1250} (\bibinfo {year} {2019})}\BibitemShut {NoStop}%
\bibitem [{\citenamefont {Knebel}\ \emph {et~al.}(2019)\citenamefont {Knebel}, \citenamefont {Knafo}, \citenamefont {Pourret}, \citenamefont {Niu}, \citenamefont {Vali{\v{s}}ka}, \citenamefont {Braithwaite}, \citenamefont {Lapertot}, \citenamefont {Nardone}, \citenamefont {Zitouni}, \citenamefont {Mishra}, \citenamefont {Sheikin}, \citenamefont {Seyfarth}, \citenamefont {Brison}, \citenamefont {Aoki},\ and\ \citenamefont {Flouquet}}]{Knebel2019}%
  \BibitemOpen
  \bibfield  {author} {\bibinfo {author} {\bibfnamefont {G.}~\bibnamefont {Knebel}}, \bibinfo {author} {\bibfnamefont {W.}~\bibnamefont {Knafo}}, \bibinfo {author} {\bibfnamefont {A.}~\bibnamefont {Pourret}}, \bibinfo {author} {\bibfnamefont {Q.}~\bibnamefont {Niu}}, \bibinfo {author} {\bibfnamefont {M.}~\bibnamefont {Vali{\v{s}}ka}}, \bibinfo {author} {\bibfnamefont {D.}~\bibnamefont {Braithwaite}}, \bibinfo {author} {\bibfnamefont {G.}~\bibnamefont {Lapertot}}, \bibinfo {author} {\bibfnamefont {M.}~\bibnamefont {Nardone}}, \bibinfo {author} {\bibfnamefont {A.}~\bibnamefont {Zitouni}}, \bibinfo {author} {\bibfnamefont {S.}~\bibnamefont {Mishra}}, \bibinfo {author} {\bibfnamefont {I.}~\bibnamefont {Sheikin}}, \bibinfo {author} {\bibfnamefont {G.}~\bibnamefont {Seyfarth}}, \bibinfo {author} {\bibfnamefont {J.~P.}\ \bibnamefont {Brison}}, \bibinfo {author} {\bibfnamefont {D.}~\bibnamefont {Aoki}}, \ and\ \bibinfo {author} {\bibfnamefont {J.}~\bibnamefont {Flouquet}},\ }\href {\doibase 10.7566/JPSJ.88.063707}
  {\bibfield  {journal} {\bibinfo  {journal} {J. Phys. Soc. Jpn.}\ }\textbf {\bibinfo {volume} {88}},\ \bibinfo {pages} {63707} (\bibinfo {year} {2019})}\BibitemShut {NoStop}%
\bibitem [{\citenamefont {Sakai}\ \emph {et~al.}(2023)\citenamefont {Sakai}, \citenamefont {Tokiwa}, \citenamefont {Opletal}, \citenamefont {Kimata}, \citenamefont {Awaji}, \citenamefont {Sasaki}, \citenamefont {Aoki}, \citenamefont {Kambe}, \citenamefont {Tokunaga},\ and\ \citenamefont {Haga}}]{Aoki_Hard}%
  \BibitemOpen
  \bibfield  {author} {\bibinfo {author} {\bibfnamefont {H.}~\bibnamefont {Sakai}}, \bibinfo {author} {\bibfnamefont {Y.}~\bibnamefont {Tokiwa}}, \bibinfo {author} {\bibfnamefont {P.}~\bibnamefont {Opletal}}, \bibinfo {author} {\bibfnamefont {M.}~\bibnamefont {Kimata}}, \bibinfo {author} {\bibfnamefont {S.}~\bibnamefont {Awaji}}, \bibinfo {author} {\bibfnamefont {T.}~\bibnamefont {Sasaki}}, \bibinfo {author} {\bibfnamefont {D.}~\bibnamefont {Aoki}}, \bibinfo {author} {\bibfnamefont {S.}~\bibnamefont {Kambe}}, \bibinfo {author} {\bibfnamefont {Y.}~\bibnamefont {Tokunaga}}, \ and\ \bibinfo {author} {\bibfnamefont {Y.}~\bibnamefont {Haga}},\ }\href {\doibase 10.1103/PhysRevLett.130.196002} {\bibfield  {journal} {\bibinfo  {journal} {Phys. Rev. Lett.}\ }\textbf {\bibinfo {volume} {130}},\ \bibinfo {pages} {196002} (\bibinfo {year} {2023})}\BibitemShut {NoStop}%
\bibitem [{\citenamefont {Wu}\ \emph {et~al.}(2024)\citenamefont {Wu}, \citenamefont {Weinberger}, \citenamefont {Chen}, \citenamefont {Cabala}, \citenamefont {Chichinadze}, \citenamefont {Shaffer}, \citenamefont {Pospíšil}, \citenamefont {Prokleška}, \citenamefont {Haidamak}, \citenamefont {Bastien}, \citenamefont {Sechovský}, \citenamefont {Hickey}, \citenamefont {Mancera-Ugarte}, \citenamefont {Benjamin}, \citenamefont {Graf}, \citenamefont {Skourski}, \citenamefont {Lonzarich}, \citenamefont {Vališka}, \citenamefont {Grosche},\ and\ \citenamefont {Eaton}}]{tony2024enhanced}%
  \BibitemOpen
  \bibfield  {author} {\bibinfo {author} {\bibfnamefont {Z.}~\bibnamefont {Wu}}, \bibinfo {author} {\bibfnamefont {T.~I.}\ \bibnamefont {Weinberger}}, \bibinfo {author} {\bibfnamefont {J.}~\bibnamefont {Chen}}, \bibinfo {author} {\bibfnamefont {A.}~\bibnamefont {Cabala}}, \bibinfo {author} {\bibfnamefont {D.~V.}\ \bibnamefont {Chichinadze}}, \bibinfo {author} {\bibfnamefont {D.}~\bibnamefont {Shaffer}}, \bibinfo {author} {\bibfnamefont {J.}~\bibnamefont {Pospíšil}}, \bibinfo {author} {\bibfnamefont {J.}~\bibnamefont {Prokleška}}, \bibinfo {author} {\bibfnamefont {T.}~\bibnamefont {Haidamak}}, \bibinfo {author} {\bibfnamefont {G.}~\bibnamefont {Bastien}}, \bibinfo {author} {\bibfnamefont {V.}~\bibnamefont {Sechovský}}, \bibinfo {author} {\bibfnamefont {A.~J.}\ \bibnamefont {Hickey}}, \bibinfo {author} {\bibfnamefont {M.~J.}\ \bibnamefont {Mancera-Ugarte}}, \bibinfo {author} {\bibfnamefont {S.}~\bibnamefont {Benjamin}}, \bibinfo {author} {\bibfnamefont {D.~E.}\ \bibnamefont {Graf}}, \bibinfo {author}
  {\bibfnamefont {Y.}~\bibnamefont {Skourski}}, \bibinfo {author} {\bibfnamefont {G.~G.}\ \bibnamefont {Lonzarich}}, \bibinfo {author} {\bibfnamefont {M.}~\bibnamefont {Vališka}}, \bibinfo {author} {\bibfnamefont {F.~M.}\ \bibnamefont {Grosche}}, \ and\ \bibinfo {author} {\bibfnamefont {A.~G.}\ \bibnamefont {Eaton}},\ }\href {https://doi.org/10.1073/pnas.2403067121} {\bibfield  {journal} {\bibinfo  {journal} {Proc. Natl. Acad. Sci. USA}\ }\textbf {\bibinfo {volume} {121}},\ \bibinfo {pages} {e2403067121} (\bibinfo {year} {2024})}\BibitemShut {NoStop}%
\bibitem [{\citenamefont {Zhang}\ \emph {et~al.}(2025)\citenamefont {Zhang}, \citenamefont {Guo}, \citenamefont {Graf}, \citenamefont {Putzke}, \citenamefont {Bordelon}, \citenamefont {Bauer}, \citenamefont {Thomas}, \citenamefont {Ronning}, \citenamefont {Rosa},\ and\ \citenamefont {Moll}}]{zhang2025dimensionality}%
  \BibitemOpen
  \bibfield  {author} {\bibinfo {author} {\bibfnamefont {L.}~\bibnamefont {Zhang}}, \bibinfo {author} {\bibfnamefont {C.}~\bibnamefont {Guo}}, \bibinfo {author} {\bibfnamefont {D.}~\bibnamefont {Graf}}, \bibinfo {author} {\bibfnamefont {C.}~\bibnamefont {Putzke}}, \bibinfo {author} {\bibfnamefont {M.}~\bibnamefont {Bordelon}}, \bibinfo {author} {\bibfnamefont {E.}~\bibnamefont {Bauer}}, \bibinfo {author} {\bibfnamefont {S.}~\bibnamefont {Thomas}}, \bibinfo {author} {\bibfnamefont {F.}~\bibnamefont {Ronning}}, \bibinfo {author} {\bibfnamefont {P.}~\bibnamefont {Rosa}}, \ and\ \bibinfo {author} {\bibfnamefont {P.}~\bibnamefont {Moll}},\ }\href {https://doi.org/10.1038/s41467-025-66288-5} {\bibfield  {journal} {\bibinfo  {journal} {Nat. Commun.}\ }\textbf {\bibinfo {volume} {16}},\ \bibinfo {pages} {10308} (\bibinfo {year} {2025})}\BibitemShut {NoStop}%
\bibitem [{\citenamefont {Wu}\ \emph {et~al.}(2026{\natexlab{a}})\citenamefont {Wu}, \citenamefont {Chen}, \citenamefont {Long}, \citenamefont {Shaffer}, \citenamefont {Chichinadze}, \citenamefont {Cabala}, \citenamefont {Weinberger}, \citenamefont {Hickey}, \citenamefont {Pu}, \citenamefont {Graf}, \citenamefont {Sechovsky}, \citenamefont {Valiska}, \citenamefont {Li}, \citenamefont {Zhou}, \citenamefont {Grosche},\ and\ \citenamefont {Eaton}}]{wu2026electricallycontrollablesuperconductingmemoryeffect}%
  \BibitemOpen
  \bibfield  {author} {\bibinfo {author} {\bibfnamefont {Z.}~\bibnamefont {Wu}}, \bibinfo {author} {\bibfnamefont {H.}~\bibnamefont {Chen}}, \bibinfo {author} {\bibfnamefont {M.}~\bibnamefont {Long}}, \bibinfo {author} {\bibfnamefont {D.}~\bibnamefont {Shaffer}}, \bibinfo {author} {\bibfnamefont {D.~V.}\ \bibnamefont {Chichinadze}}, \bibinfo {author} {\bibfnamefont {A.}~\bibnamefont {Cabala}}, \bibinfo {author} {\bibfnamefont {T.~I.}\ \bibnamefont {Weinberger}}, \bibinfo {author} {\bibfnamefont {A.~J.}\ \bibnamefont {Hickey}}, \bibinfo {author} {\bibfnamefont {J.}~\bibnamefont {Pu}}, \bibinfo {author} {\bibfnamefont {D.}~\bibnamefont {Graf}}, \bibinfo {author} {\bibfnamefont {V.}~\bibnamefont {Sechovsky}}, \bibinfo {author} {\bibfnamefont {M.}~\bibnamefont {Valiska}}, \bibinfo {author} {\bibfnamefont {G.}~\bibnamefont {Li}}, \bibinfo {author} {\bibfnamefont {R.}~\bibnamefont {Zhou}}, \bibinfo {author} {\bibfnamefont {F.~M.}\ \bibnamefont {Grosche}}, \ and\ \bibinfo {author} {\bibfnamefont {A.~G.}\
  \bibnamefont {Eaton}},\ }\href {https://arxiv.org/abs/2603.02450} {\enquote {\bibinfo {title} {{Electrically-controllable superconducting memory effect in UTe$_2$}},}\ } (\bibinfo {year} {2026}{\natexlab{a}}),\ \Eprint {http://arxiv.org/abs/2603.02450} {arXiv:2603.02450 [cond-mat.supr-con]} \BibitemShut {NoStop}%
\bibitem [{\citenamefont {Knafo}\ \emph {et~al.}(2021)\citenamefont {Knafo}, \citenamefont {Nardone}, \citenamefont {Vali{\v{s}}ka}, \citenamefont {Zitouni}, \citenamefont {Lapertot}, \citenamefont {Aoki}, \citenamefont {Knebel},\ and\ \citenamefont {Braithwaite}}]{knafo2021comparison}%
  \BibitemOpen
  \bibfield  {author} {\bibinfo {author} {\bibfnamefont {W.}~\bibnamefont {Knafo}}, \bibinfo {author} {\bibfnamefont {M.}~\bibnamefont {Nardone}}, \bibinfo {author} {\bibfnamefont {M.}~\bibnamefont {Vali{\v{s}}ka}}, \bibinfo {author} {\bibfnamefont {A.}~\bibnamefont {Zitouni}}, \bibinfo {author} {\bibfnamefont {G.}~\bibnamefont {Lapertot}}, \bibinfo {author} {\bibfnamefont {D.}~\bibnamefont {Aoki}}, \bibinfo {author} {\bibfnamefont {G.}~\bibnamefont {Knebel}}, \ and\ \bibinfo {author} {\bibfnamefont {D.}~\bibnamefont {Braithwaite}},\ }\href {https://doi.org/10.1038/s42005-021-00545-z} {\bibfield  {journal} {\bibinfo  {journal} {Commun. Phys.}\ }\textbf {\bibinfo {volume} {4}},\ \bibinfo {pages} {40} (\bibinfo {year} {2021})}\BibitemShut {NoStop}%
\bibitem [{\citenamefont {Helm}\ \emph {et~al.}(2024)\citenamefont {Helm}, \citenamefont {Kimata}, \citenamefont {Sudo}, \citenamefont {Miyata}, \citenamefont {Stirnat}, \citenamefont {F{\"o}rster}, \citenamefont {Hornung}, \citenamefont {K{\"o}nig}, \citenamefont {Sheikin}, \citenamefont {Pourret} \emph {et~al.}}]{helm2024}%
  \BibitemOpen
  \bibfield  {author} {\bibinfo {author} {\bibfnamefont {T.}~\bibnamefont {Helm}}, \bibinfo {author} {\bibfnamefont {M.}~\bibnamefont {Kimata}}, \bibinfo {author} {\bibfnamefont {K.}~\bibnamefont {Sudo}}, \bibinfo {author} {\bibfnamefont {A.}~\bibnamefont {Miyata}}, \bibinfo {author} {\bibfnamefont {J.}~\bibnamefont {Stirnat}}, \bibinfo {author} {\bibfnamefont {T.}~\bibnamefont {F{\"o}rster}}, \bibinfo {author} {\bibfnamefont {J.}~\bibnamefont {Hornung}}, \bibinfo {author} {\bibfnamefont {M.}~\bibnamefont {K{\"o}nig}}, \bibinfo {author} {\bibfnamefont {I.}~\bibnamefont {Sheikin}}, \bibinfo {author} {\bibfnamefont {A.}~\bibnamefont {Pourret}},  \emph {et~al.},\ }\href {https://doi.org/10.1038/s41467-023-44183-1} {\bibfield  {journal} {\bibinfo  {journal} {Nat. Commun.}\ }\textbf {\bibinfo {volume} {15}},\ \bibinfo {pages} {37} (\bibinfo {year} {2024})}\BibitemShut {NoStop}%
\bibitem [{\citenamefont {Sch{\"o}nemann}\ \emph {et~al.}(2024)\citenamefont {Sch{\"o}nemann}, \citenamefont {Rosa}, \citenamefont {Thomas}, \citenamefont {Lai}, \citenamefont {Nguyen}, \citenamefont {Singleton}, \citenamefont {Brosha}, \citenamefont {McDonald}, \citenamefont {Zapf}, \citenamefont {Maiorov} \emph {et~al.}}]{LANL_bulk_UTe2}%
  \BibitemOpen
  \bibfield  {author} {\bibinfo {author} {\bibfnamefont {R.}~\bibnamefont {Sch{\"o}nemann}}, \bibinfo {author} {\bibfnamefont {P.~F.}\ \bibnamefont {Rosa}}, \bibinfo {author} {\bibfnamefont {S.~M.}\ \bibnamefont {Thomas}}, \bibinfo {author} {\bibfnamefont {Y.}~\bibnamefont {Lai}}, \bibinfo {author} {\bibfnamefont {D.~N.}\ \bibnamefont {Nguyen}}, \bibinfo {author} {\bibfnamefont {J.}~\bibnamefont {Singleton}}, \bibinfo {author} {\bibfnamefont {E.~L.}\ \bibnamefont {Brosha}}, \bibinfo {author} {\bibfnamefont {R.~D.}\ \bibnamefont {McDonald}}, \bibinfo {author} {\bibfnamefont {V.}~\bibnamefont {Zapf}}, \bibinfo {author} {\bibfnamefont {B.}~\bibnamefont {Maiorov}},  \emph {et~al.},\ }\href {https://doi.org/10.1093/pnasnexus/pgad428} {\bibfield  {journal} {\bibinfo  {journal} {PNAS Nexus}\ }\textbf {\bibinfo {volume} {3}},\ \bibinfo {pages} {pgad428} (\bibinfo {year} {2024})}\BibitemShut {NoStop}%
\bibitem [{\citenamefont {Frank}\ \emph {et~al.}(2024)\citenamefont {Frank}, \citenamefont {Lewin}, \citenamefont {Saucedo~Salas}, \citenamefont {Czajka}, \citenamefont {Hayes}, \citenamefont {Yoon}, \citenamefont {Metz}, \citenamefont {Paglione}, \citenamefont {Singleton},\ and\ \citenamefont {Butch}}]{frank2024orphan}%
  \BibitemOpen
  \bibfield  {author} {\bibinfo {author} {\bibfnamefont {C.~E.}\ \bibnamefont {Frank}}, \bibinfo {author} {\bibfnamefont {S.~K.}\ \bibnamefont {Lewin}}, \bibinfo {author} {\bibfnamefont {G.}~\bibnamefont {Saucedo~Salas}}, \bibinfo {author} {\bibfnamefont {P.}~\bibnamefont {Czajka}}, \bibinfo {author} {\bibfnamefont {I.~M.}\ \bibnamefont {Hayes}}, \bibinfo {author} {\bibfnamefont {H.}~\bibnamefont {Yoon}}, \bibinfo {author} {\bibfnamefont {T.}~\bibnamefont {Metz}}, \bibinfo {author} {\bibfnamefont {J.}~\bibnamefont {Paglione}}, \bibinfo {author} {\bibfnamefont {J.}~\bibnamefont {Singleton}}, \ and\ \bibinfo {author} {\bibfnamefont {N.~P.}\ \bibnamefont {Butch}},\ }\href {https://doi.org/10.1038/s41467-024-47090-1} {\bibfield  {journal} {\bibinfo  {journal} {Nat. Commun.}\ }\textbf {\bibinfo {volume} {15}},\ \bibinfo {pages} {3378} (\bibinfo {year} {2024})}\BibitemShut {NoStop}%
\bibitem [{\citenamefont {Wu}\ \emph {et~al.}(2025{\natexlab{a}})\citenamefont {Wu}, \citenamefont {Chen}, \citenamefont {Weinberger}, \citenamefont {Cabala}, \citenamefont {Graf}, \citenamefont {Skourski}, \citenamefont {Xie}, \citenamefont {Ling}, \citenamefont {Zhu}, \citenamefont {Sechovský}, \citenamefont {Vališka}, \citenamefont {Grosche},\ and\ \citenamefont {Eaton}}]{tony2025brief}%
  \BibitemOpen
  \bibfield  {author} {\bibinfo {author} {\bibfnamefont {Z.}~\bibnamefont {Wu}}, \bibinfo {author} {\bibfnamefont {H.}~\bibnamefont {Chen}}, \bibinfo {author} {\bibfnamefont {T.~I.}\ \bibnamefont {Weinberger}}, \bibinfo {author} {\bibfnamefont {A.}~\bibnamefont {Cabala}}, \bibinfo {author} {\bibfnamefont {D.~E.}\ \bibnamefont {Graf}}, \bibinfo {author} {\bibfnamefont {Y.}~\bibnamefont {Skourski}}, \bibinfo {author} {\bibfnamefont {W.}~\bibnamefont {Xie}}, \bibinfo {author} {\bibfnamefont {Y.}~\bibnamefont {Ling}}, \bibinfo {author} {\bibfnamefont {Z.}~\bibnamefont {Zhu}}, \bibinfo {author} {\bibfnamefont {V.}~\bibnamefont {Sechovský}}, \bibinfo {author} {\bibfnamefont {M.}~\bibnamefont {Vališka}}, \bibinfo {author} {\bibfnamefont {F.~M.}\ \bibnamefont {Grosche}}, \ and\ \bibinfo {author} {\bibfnamefont {A.~G.}\ \bibnamefont {Eaton}},\ }\href {https://doi.org/10.1073/pnas.2422156122} {\bibfield  {journal} {\bibinfo  {journal} {Proc. Natl. Acad. Sci. USA}\ }\textbf {\bibinfo {volume} {122}},\ \bibinfo {pages}
  {e2422156122} (\bibinfo {year} {2025}{\natexlab{a}})}\BibitemShut {NoStop}%
\bibitem [{\citenamefont {Wu}\ \emph {et~al.}(2025{\natexlab{b}})\citenamefont {Wu}, \citenamefont {Weinberger}, \citenamefont {Hickey}, \citenamefont {Chichinadze}, \citenamefont {Shaffer}, \citenamefont {Cabala}, \citenamefont {Chen}, \citenamefont {Long}, \citenamefont {Brumm}, \citenamefont {Xie}, \citenamefont {Ling}, \citenamefont {Zhu}, \citenamefont {Skourski}, \citenamefont {Graf}, \citenamefont {Sechovsk\'y}, \citenamefont {Vali\ifmmode~\check{s}\else \v{s}\fi{}ka}, \citenamefont {Lonzarich}, \citenamefont {Grosche},\ and\ \citenamefont {Eaton}}]{qcl}%
  \BibitemOpen
  \bibfield  {author} {\bibinfo {author} {\bibfnamefont {Z.}~\bibnamefont {Wu}}, \bibinfo {author} {\bibfnamefont {T.~I.}\ \bibnamefont {Weinberger}}, \bibinfo {author} {\bibfnamefont {A.~J.}\ \bibnamefont {Hickey}}, \bibinfo {author} {\bibfnamefont {D.~V.}\ \bibnamefont {Chichinadze}}, \bibinfo {author} {\bibfnamefont {D.}~\bibnamefont {Shaffer}}, \bibinfo {author} {\bibfnamefont {A.}~\bibnamefont {Cabala}}, \bibinfo {author} {\bibfnamefont {H.}~\bibnamefont {Chen}}, \bibinfo {author} {\bibfnamefont {M.}~\bibnamefont {Long}}, \bibinfo {author} {\bibfnamefont {T.~J.}\ \bibnamefont {Brumm}}, \bibinfo {author} {\bibfnamefont {W.}~\bibnamefont {Xie}}, \bibinfo {author} {\bibfnamefont {Y.}~\bibnamefont {Ling}}, \bibinfo {author} {\bibfnamefont {Z.}~\bibnamefont {Zhu}}, \bibinfo {author} {\bibfnamefont {Y.}~\bibnamefont {Skourski}}, \bibinfo {author} {\bibfnamefont {D.~E.}\ \bibnamefont {Graf}}, \bibinfo {author} {\bibfnamefont {V.}~\bibnamefont {Sechovsk\'y}}, \bibinfo {author} {\bibfnamefont {M.}~\bibnamefont
  {Vali\ifmmode~\check{s}\else \v{s}\fi{}ka}}, \bibinfo {author} {\bibfnamefont {G.~G.}\ \bibnamefont {Lonzarich}}, \bibinfo {author} {\bibfnamefont {F.~M.}\ \bibnamefont {Grosche}}, \ and\ \bibinfo {author} {\bibfnamefont {A.~G.}\ \bibnamefont {Eaton}},\ }\href {\doibase 10.1103/PhysRevX.15.021019} {\bibfield  {journal} {\bibinfo  {journal} {Phys. Rev. X}\ }\textbf {\bibinfo {volume} {15}},\ \bibinfo {pages} {021019} (\bibinfo {year} {2025}{\natexlab{b}})}\BibitemShut {NoStop}%
\bibitem [{\citenamefont {Lewin}\ \emph {et~al.}(2025)\citenamefont {Lewin}, \citenamefont {Czajka}, \citenamefont {Frank}, \citenamefont {Saucedo~Salas}, \citenamefont {Noe~II}, \citenamefont {Yoon}, \citenamefont {Eo}, \citenamefont {Paglione}, \citenamefont {Nevidomskyy}, \citenamefont {Singleton},\ and\ \citenamefont {Butch}}]{lewin2025halo}%
  \BibitemOpen
  \bibfield  {author} {\bibinfo {author} {\bibfnamefont {S.~K.}\ \bibnamefont {Lewin}}, \bibinfo {author} {\bibfnamefont {P.}~\bibnamefont {Czajka}}, \bibinfo {author} {\bibfnamefont {C.~E.}\ \bibnamefont {Frank}}, \bibinfo {author} {\bibfnamefont {G.}~\bibnamefont {Saucedo~Salas}}, \bibinfo {author} {\bibfnamefont {G.~T.}\ \bibnamefont {Noe~II}}, \bibinfo {author} {\bibfnamefont {H.}~\bibnamefont {Yoon}}, \bibinfo {author} {\bibfnamefont {Y.~S.}\ \bibnamefont {Eo}}, \bibinfo {author} {\bibfnamefont {J.}~\bibnamefont {Paglione}}, \bibinfo {author} {\bibfnamefont {A.~H.}\ \bibnamefont {Nevidomskyy}}, \bibinfo {author} {\bibfnamefont {J.}~\bibnamefont {Singleton}}, \ and\ \bibinfo {author} {\bibfnamefont {N.~P.}\ \bibnamefont {Butch}},\ }\href {https://doi.org/10.1126/science.adn7673} {\bibfield  {journal} {\bibinfo  {journal} {Science}\ }\textbf {\bibinfo {volume} {389}},\ \bibinfo {pages} {512} (\bibinfo {year} {2025})}\BibitemShut {NoStop}%
\bibitem [{\citenamefont {Weinberger}\ \emph {et~al.}(2025)\citenamefont {Weinberger}, \citenamefont {Chen}, \citenamefont {Wu}, \citenamefont {Long}, \citenamefont {Cabala}, \citenamefont {Skourski}, \citenamefont {Sourd}, \citenamefont {Haidamak}, \citenamefont {Sechovsky}, \citenamefont {Valiska}, \citenamefont {Grosche},\ and\ \citenamefont {Eaton}}]{weinberger2025strangemetallicityencompasseshigh}%
  \BibitemOpen
  \bibfield  {author} {\bibinfo {author} {\bibfnamefont {T.~I.}\ \bibnamefont {Weinberger}}, \bibinfo {author} {\bibfnamefont {H.}~\bibnamefont {Chen}}, \bibinfo {author} {\bibfnamefont {Z.}~\bibnamefont {Wu}}, \bibinfo {author} {\bibfnamefont {M.}~\bibnamefont {Long}}, \bibinfo {author} {\bibfnamefont {A.}~\bibnamefont {Cabala}}, \bibinfo {author} {\bibfnamefont {Y.}~\bibnamefont {Skourski}}, \bibinfo {author} {\bibfnamefont {J.}~\bibnamefont {Sourd}}, \bibinfo {author} {\bibfnamefont {T.}~\bibnamefont {Haidamak}}, \bibinfo {author} {\bibfnamefont {V.}~\bibnamefont {Sechovsky}}, \bibinfo {author} {\bibfnamefont {M.}~\bibnamefont {Valiska}}, \bibinfo {author} {\bibfnamefont {F.~M.}\ \bibnamefont {Grosche}}, \ and\ \bibinfo {author} {\bibfnamefont {A.~G.}\ \bibnamefont {Eaton}},\ }\href {https://arxiv.org/abs/2505.12131} {\enquote {\bibinfo {title} {{Strange metallicity encompasses high magnetic field-induced superconductivity in UTe$_2$}},}\ } (\bibinfo {year} {2025}),\ \Eprint
  {http://arxiv.org/abs/2505.12131} {arXiv:2505.12131 [cond-mat.str-el]} \BibitemShut {NoStop}%
\bibitem [{\citenamefont {Wu}\ \emph {et~al.}(2026{\natexlab{b}})\citenamefont {Wu}, \citenamefont {Chen}, \citenamefont {Weinberger}, \citenamefont {Long}, \citenamefont {Graf}, \citenamefont {Cabala}, \citenamefont {Sechovsky}, \citenamefont {Valiska}, \citenamefont {Lonzarich}, \citenamefont {Grosche},\ and\ \citenamefont {Eaton}}]{wu2026directobservationspilloverhigh}%
  \BibitemOpen
  \bibfield  {author} {\bibinfo {author} {\bibfnamefont {Z.}~\bibnamefont {Wu}}, \bibinfo {author} {\bibfnamefont {H.}~\bibnamefont {Chen}}, \bibinfo {author} {\bibfnamefont {T.~I.}\ \bibnamefont {Weinberger}}, \bibinfo {author} {\bibfnamefont {M.}~\bibnamefont {Long}}, \bibinfo {author} {\bibfnamefont {D.}~\bibnamefont {Graf}}, \bibinfo {author} {\bibfnamefont {A.}~\bibnamefont {Cabala}}, \bibinfo {author} {\bibfnamefont {V.}~\bibnamefont {Sechovsky}}, \bibinfo {author} {\bibfnamefont {M.}~\bibnamefont {Valiska}}, \bibinfo {author} {\bibfnamefont {G.~G.}\ \bibnamefont {Lonzarich}}, \bibinfo {author} {\bibfnamefont {F.~M.}\ \bibnamefont {Grosche}}, \ and\ \bibinfo {author} {\bibfnamefont {A.~G.}\ \bibnamefont {Eaton}},\ }\href {https://arxiv.org/abs/2601.04594} {\enquote {\bibinfo {title} {{Direct Observation of the Spillover of High Magnetic Field-induced SC3 Superconductivity Outside the Spin-Polarized State in UTe2}},}\ } (\bibinfo {year} {2026}{\natexlab{b}}),\ \Eprint {http://arxiv.org/abs/2601.04594}
  {arXiv:2601.04594 [cond-mat.supr-con]} \BibitemShut {NoStop}%
\bibitem [{\citenamefont {Tokunaga}\ \emph {et~al.}(2023)\citenamefont {Tokunaga}, \citenamefont {Sakai}, \citenamefont {Kambe}, \citenamefont {Opletal}, \citenamefont {Tokiwa}, \citenamefont {Haga}, \citenamefont {Kitagawa}, \citenamefont {Ishida}, \citenamefont {Aoki}, \citenamefont {Knebel}, \citenamefont {Lapertot}, \citenamefont {Kr\"amer},\ and\ \citenamefont {Horvati\ifmmode~\acute{c}\else \'{c}\fi{}}}]{tokunaga2023longitudinal}%
  \BibitemOpen
  \bibfield  {author} {\bibinfo {author} {\bibfnamefont {Y.}~\bibnamefont {Tokunaga}}, \bibinfo {author} {\bibfnamefont {H.}~\bibnamefont {Sakai}}, \bibinfo {author} {\bibfnamefont {S.}~\bibnamefont {Kambe}}, \bibinfo {author} {\bibfnamefont {P.}~\bibnamefont {Opletal}}, \bibinfo {author} {\bibfnamefont {Y.}~\bibnamefont {Tokiwa}}, \bibinfo {author} {\bibfnamefont {Y.}~\bibnamefont {Haga}}, \bibinfo {author} {\bibfnamefont {S.}~\bibnamefont {Kitagawa}}, \bibinfo {author} {\bibfnamefont {K.}~\bibnamefont {Ishida}}, \bibinfo {author} {\bibfnamefont {D.}~\bibnamefont {Aoki}}, \bibinfo {author} {\bibfnamefont {G.}~\bibnamefont {Knebel}}, \bibinfo {author} {\bibfnamefont {G.}~\bibnamefont {Lapertot}}, \bibinfo {author} {\bibfnamefont {S.}~\bibnamefont {Kr\"amer}}, \ and\ \bibinfo {author} {\bibfnamefont {M.}~\bibnamefont {Horvati\ifmmode~\acute{c}\else \'{c}\fi{}}},\ }\href {\doibase 10.1103/PhysRevLett.131.226503} {\bibfield  {journal} {\bibinfo  {journal} {Phys. Rev. Lett.}\ }\textbf {\bibinfo {volume} {131}},\
  \bibinfo {pages} {226503} (\bibinfo {year} {2023})}\BibitemShut {NoStop}%
\bibitem [{\citenamefont {Tokiwa}\ \emph {et~al.}(2024)\citenamefont {Tokiwa}, \citenamefont {Opletal}, \citenamefont {Sakai}, \citenamefont {Kambe}, \citenamefont {Yamamoto}, \citenamefont {Kimata}, \citenamefont {Awaji}, \citenamefont {Sasaki}, \citenamefont {Aoki}, \citenamefont {Haga},\ and\ \citenamefont {Tokunaga}}]{TokiwaPRB2024}%
  \BibitemOpen
  \bibfield  {author} {\bibinfo {author} {\bibfnamefont {Y.}~\bibnamefont {Tokiwa}}, \bibinfo {author} {\bibfnamefont {P.}~\bibnamefont {Opletal}}, \bibinfo {author} {\bibfnamefont {H.}~\bibnamefont {Sakai}}, \bibinfo {author} {\bibfnamefont {S.}~\bibnamefont {Kambe}}, \bibinfo {author} {\bibfnamefont {E.}~\bibnamefont {Yamamoto}}, \bibinfo {author} {\bibfnamefont {M.}~\bibnamefont {Kimata}}, \bibinfo {author} {\bibfnamefont {S.}~\bibnamefont {Awaji}}, \bibinfo {author} {\bibfnamefont {T.}~\bibnamefont {Sasaki}}, \bibinfo {author} {\bibfnamefont {D.}~\bibnamefont {Aoki}}, \bibinfo {author} {\bibfnamefont {Y.}~\bibnamefont {Haga}}, \ and\ \bibinfo {author} {\bibfnamefont {Y.}~\bibnamefont {Tokunaga}},\ }\href {\doibase 10.1103/PhysRevB.109.L140502} {\bibfield  {journal} {\bibinfo  {journal} {Phys. Rev. B}\ }\textbf {\bibinfo {volume} {109}},\ \bibinfo {pages} {L140502} (\bibinfo {year} {2024})}\BibitemShut {NoStop}%
\bibitem [{\citenamefont {Wu}\ \emph {et~al.}(2025{\natexlab{c}})\citenamefont {Wu}, \citenamefont {Chen}, \citenamefont {Long}, \citenamefont {Jin}, \citenamefont {Zuo}, \citenamefont {Shaffer}, \citenamefont {Chichinadze}, \citenamefont {Cabala}, \citenamefont {Sechovsky}, \citenamefont {Valiska}, \citenamefont {Zhu}, \citenamefont {Lonzarich}, \citenamefont {Grosche},\ and\ \citenamefont {Eaton}}]{ripples}%
  \BibitemOpen
  \bibfield  {author} {\bibinfo {author} {\bibfnamefont {Z.}~\bibnamefont {Wu}}, \bibinfo {author} {\bibfnamefont {H.}~\bibnamefont {Chen}}, \bibinfo {author} {\bibfnamefont {M.}~\bibnamefont {Long}}, \bibinfo {author} {\bibfnamefont {G.}~\bibnamefont {Jin}}, \bibinfo {author} {\bibfnamefont {H.}~\bibnamefont {Zuo}}, \bibinfo {author} {\bibfnamefont {D.}~\bibnamefont {Shaffer}}, \bibinfo {author} {\bibfnamefont {D.~V.}\ \bibnamefont {Chichinadze}}, \bibinfo {author} {\bibfnamefont {A.}~\bibnamefont {Cabala}}, \bibinfo {author} {\bibfnamefont {V.}~\bibnamefont {Sechovsky}}, \bibinfo {author} {\bibfnamefont {M.}~\bibnamefont {Valiska}}, \bibinfo {author} {\bibfnamefont {Z.}~\bibnamefont {Zhu}}, \bibinfo {author} {\bibfnamefont {G.~G.}\ \bibnamefont {Lonzarich}}, \bibinfo {author} {\bibfnamefont {F.~M.}\ \bibnamefont {Grosche}}, \ and\ \bibinfo {author} {\bibfnamefont {A.~G.}\ \bibnamefont {Eaton}},\ }\href {https://arxiv.org/abs/2503.11362} {\enquote {\bibinfo {title} {{Metamagnetic ripples in the UTe2 high
  magnetic field phase diagram}},}\ } (\bibinfo {year} {2025}{\natexlab{c}}),\ \Eprint {http://arxiv.org/abs/2503.11362} {arXiv:2503.11362 [cond-mat.str-el]} \BibitemShut {NoStop}%
\bibitem [{\citenamefont {Weinberger}\ \emph {et~al.}(2026)\citenamefont {Weinberger}, \citenamefont {Shaffer}, \citenamefont {Wu}, \citenamefont {Chichinadze}, \citenamefont {Pu}, \citenamefont {Li}, \citenamefont {Zhou}, \citenamefont {Skourski}, \citenamefont {Graf}, \citenamefont {Cabala}, \citenamefont {Sechovsky}, \citenamefont {Valiska}, \citenamefont {Kwasigroch}, \citenamefont {Grosche},\ and\ \citenamefont {Eaton}}]{weinberger2026metamagnetismute2rolesitinerancy}%
  \BibitemOpen
  \bibfield  {author} {\bibinfo {author} {\bibfnamefont {T.~I.}\ \bibnamefont {Weinberger}}, \bibinfo {author} {\bibfnamefont {D.}~\bibnamefont {Shaffer}}, \bibinfo {author} {\bibfnamefont {Z.}~\bibnamefont {Wu}}, \bibinfo {author} {\bibfnamefont {D.~V.}\ \bibnamefont {Chichinadze}}, \bibinfo {author} {\bibfnamefont {J.}~\bibnamefont {Pu}}, \bibinfo {author} {\bibfnamefont {G.}~\bibnamefont {Li}}, \bibinfo {author} {\bibfnamefont {R.}~\bibnamefont {Zhou}}, \bibinfo {author} {\bibfnamefont {Y.}~\bibnamefont {Skourski}}, \bibinfo {author} {\bibfnamefont {D.}~\bibnamefont {Graf}}, \bibinfo {author} {\bibfnamefont {A.}~\bibnamefont {Cabala}}, \bibinfo {author} {\bibfnamefont {V.}~\bibnamefont {Sechovsky}}, \bibinfo {author} {\bibfnamefont {M.}~\bibnamefont {Valiska}}, \bibinfo {author} {\bibfnamefont {M.~P.}\ \bibnamefont {Kwasigroch}}, \bibinfo {author} {\bibfnamefont {F.~M.}\ \bibnamefont {Grosche}}, \ and\ \bibinfo {author} {\bibfnamefont {A.~G.}\ \bibnamefont {Eaton}},\ }\href
  {https://arxiv.org/abs/2606.27913} {\enquote {\bibinfo {title} {Metamagnetism in {UTe$_2$}: the roles of itinerancy and localization},}\ } (\bibinfo {year} {2026}),\ \Eprint {http://arxiv.org/abs/2606.27913} {arXiv:2606.27913 [cond-mat.str-el]} \BibitemShut {NoStop}%
\bibitem [{\citenamefont {Wohlfarth}\ and\ \citenamefont {Rhodes}(1962)}]{Wohlfarth_Rhodes}%
  \BibitemOpen
  \bibfield  {author} {\bibinfo {author} {\bibfnamefont {E.~P.}\ \bibnamefont {Wohlfarth}}\ and\ \bibinfo {author} {\bibfnamefont {P.}~\bibnamefont {Rhodes}},\ }\href {\doibase 10.1080/14786436208213848} {\bibfield  {journal} {\bibinfo  {journal} {Philos. Mag.}\ }\textbf {\bibinfo {volume} {7}},\ \bibinfo {pages} {1817} (\bibinfo {year} {1962})}\BibitemShut {NoStop}%
\bibitem [{\citenamefont {Shimizu}(1982)}]{Shimizu_1982}%
  \BibitemOpen
  \bibfield  {author} {\bibinfo {author} {\bibfnamefont {M.}~\bibnamefont {Shimizu}},\ }\href {\doibase 10.1051/jphys:01982004301015500} {\bibfield  {journal} {\bibinfo  {journal} {J. Phys. France}\ }\textbf {\bibinfo {volume} {43}},\ \bibinfo {pages} {155} (\bibinfo {year} {1982})}\BibitemShut {NoStop}%
\bibitem [{\citenamefont {Yamada}(1993)}]{Yamada_1993}%
  \BibitemOpen
  \bibfield  {author} {\bibinfo {author} {\bibfnamefont {H.}~\bibnamefont {Yamada}},\ }\href {\doibase 10.1103/PhysRevB.47.11211} {\bibfield  {journal} {\bibinfo  {journal} {Phys. Rev. B}\ }\textbf {\bibinfo {volume} {47}},\ \bibinfo {pages} {11211} (\bibinfo {year} {1993})}\BibitemShut {NoStop}%
\bibitem [{\citenamefont {Miyake}\ and\ \citenamefont {Kuramoto}(1991)}]{MIYAKE_1991}%
  \BibitemOpen
  \bibfield  {author} {\bibinfo {author} {\bibfnamefont {K.}~\bibnamefont {Miyake}}\ and\ \bibinfo {author} {\bibfnamefont {Y.}~\bibnamefont {Kuramoto}},\ }\href {\doibase https://doi.org/10.1016/0921-4526(91)90486-X} {\bibfield  {journal} {\bibinfo  {journal} {Physica B: Condensed Matter}\ }\textbf {\bibinfo {volume} {171}},\ \bibinfo {pages} {20} (\bibinfo {year} {1991})}\BibitemShut {NoStop}%
\bibitem [{\citenamefont {Ōno}(1998)}]{Ono_1998}%
  \BibitemOpen
  \bibfield  {author} {\bibinfo {author} {\bibfnamefont {Y.}~\bibnamefont {Ōno}},\ }\href {\doibase 10.1143/JPSJ.67.2197} {\bibfield  {journal} {\bibinfo  {journal} {Journal of the Physical Society of Japan}\ }\textbf {\bibinfo {volume} {67}},\ \bibinfo {pages} {2197} (\bibinfo {year} {1998})}\BibitemShut {NoStop}%
\bibitem [{\citenamefont {Viola~Kusminskiy}\ \emph {et~al.}(2008)\citenamefont {Viola~Kusminskiy}, \citenamefont {Beach}, \citenamefont {Castro~Neto},\ and\ \citenamefont {Campbell}}]{Kusminskiy_2008}%
  \BibitemOpen
  \bibfield  {author} {\bibinfo {author} {\bibfnamefont {S.}~\bibnamefont {Viola~Kusminskiy}}, \bibinfo {author} {\bibfnamefont {K.~S.~D.}\ \bibnamefont {Beach}}, \bibinfo {author} {\bibfnamefont {A.~H.}\ \bibnamefont {Castro~Neto}}, \ and\ \bibinfo {author} {\bibfnamefont {D.~K.}\ \bibnamefont {Campbell}},\ }\href {\doibase 10.1103/PhysRevB.77.094419} {\bibfield  {journal} {\bibinfo  {journal} {Phys. Rev. B}\ }\textbf {\bibinfo {volume} {77}},\ \bibinfo {pages} {094419} (\bibinfo {year} {2008})}\BibitemShut {NoStop}%
\bibitem [{\citenamefont {Thomas}\ \emph {et~al.}(2023)\citenamefont {Thomas}, \citenamefont {Burdin},\ and\ \citenamefont {Lacroix}}]{Lacroix_5f2}%
  \BibitemOpen
  \bibfield  {author} {\bibinfo {author} {\bibfnamefont {C.}~\bibnamefont {Thomas}}, \bibinfo {author} {\bibfnamefont {S.}~\bibnamefont {Burdin}}, \ and\ \bibinfo {author} {\bibfnamefont {C.}~\bibnamefont {Lacroix}},\ }\href {\doibase 10.1088/1361-648x/ace57a} {\bibfield  {journal} {\bibinfo  {journal} {Journal of Physics Condensed Matter}\ }\textbf {\bibinfo {volume} {35}},\ \bibinfo {pages} {445601} (\bibinfo {year} {2023})}\BibitemShut {NoStop}%
\bibitem [{\citenamefont {Aoki}\ \emph {et~al.}(2013)\citenamefont {Aoki}, \citenamefont {Knafo},\ and\ \citenamefont {Sheikin}}]{Aoki_2013}%
  \BibitemOpen
  \bibfield  {author} {\bibinfo {author} {\bibfnamefont {D.}~\bibnamefont {Aoki}}, \bibinfo {author} {\bibfnamefont {W.}~\bibnamefont {Knafo}}, \ and\ \bibinfo {author} {\bibfnamefont {I.}~\bibnamefont {Sheikin}},\ }\href {\doibase https://doi.org/10.1016/j.crhy.2012.11.004} {\bibfield  {journal} {\bibinfo  {journal} {Comptes Rendus Physique}\ }\textbf {\bibinfo {volume} {14}},\ \bibinfo {pages} {53} (\bibinfo {year} {2013})}\BibitemShut {NoStop}%
\bibitem [{\citenamefont {Goto}\ \emph {et~al.}(1994)\citenamefont {Goto}, \citenamefont {Katori}, \citenamefont {Sakakibara}, \citenamefont {Mitamura}, \citenamefont {Fukamichi},\ and\ \citenamefont {Murata}}]{Goto_1994}%
  \BibitemOpen
  \bibfield  {author} {\bibinfo {author} {\bibfnamefont {T.}~\bibnamefont {Goto}}, \bibinfo {author} {\bibfnamefont {H.~A.}\ \bibnamefont {Katori}}, \bibinfo {author} {\bibfnamefont {T.}~\bibnamefont {Sakakibara}}, \bibinfo {author} {\bibfnamefont {H.}~\bibnamefont {Mitamura}}, \bibinfo {author} {\bibfnamefont {K.}~\bibnamefont {Fukamichi}}, \ and\ \bibinfo {author} {\bibfnamefont {K.}~\bibnamefont {Murata}},\ }\href {\doibase 10.1063/1.358167} {\bibfield  {journal} {\bibinfo  {journal} {Journal of Applied Physics}\ }\textbf {\bibinfo {volume} {76}},\ \bibinfo {pages} {6682} (\bibinfo {year} {1994})}\BibitemShut {NoStop}%
\bibitem [{\citenamefont {Curro}\ \emph {et~al.}(2004)\citenamefont {Curro}, \citenamefont {Young}, \citenamefont {Schmalian},\ and\ \citenamefont {Pines}}]{Schmallian}%
  \BibitemOpen
  \bibfield  {author} {\bibinfo {author} {\bibfnamefont {N.~J.}\ \bibnamefont {Curro}}, \bibinfo {author} {\bibfnamefont {B.-L.}\ \bibnamefont {Young}}, \bibinfo {author} {\bibfnamefont {J.}~\bibnamefont {Schmalian}}, \ and\ \bibinfo {author} {\bibfnamefont {D.}~\bibnamefont {Pines}},\ }\href {\doibase 10.1103/PhysRevB.70.235117} {\bibfield  {journal} {\bibinfo  {journal} {Phys. Rev. B}\ }\textbf {\bibinfo {volume} {70}},\ \bibinfo {pages} {235117} (\bibinfo {year} {2004})}\BibitemShut {NoStop}%
\bibitem [{\citenamefont {Jiang}\ \emph {et~al.}(2014)\citenamefont {Jiang}, \citenamefont {Curro},\ and\ \citenamefont {Scalettar}}]{Curro_QM}%
  \BibitemOpen
  \bibfield  {author} {\bibinfo {author} {\bibfnamefont {M.}~\bibnamefont {Jiang}}, \bibinfo {author} {\bibfnamefont {N.~J.}\ \bibnamefont {Curro}}, \ and\ \bibinfo {author} {\bibfnamefont {R.~T.}\ \bibnamefont {Scalettar}},\ }\href {\doibase 10.1103/PhysRevB.90.241109} {\bibfield  {journal} {\bibinfo  {journal} {Phys. Rev. B}\ }\textbf {\bibinfo {volume} {90}},\ \bibinfo {pages} {241109} (\bibinfo {year} {2014})}\BibitemShut {NoStop}%
\bibitem [{\citenamefont {Curro}(2016)}]{Curro_review}%
  \BibitemOpen
  \bibfield  {author} {\bibinfo {author} {\bibfnamefont {N.~J.}\ \bibnamefont {Curro}},\ }\href {\doibase 10.1088/0034-4885/79/6/064501} {\bibfield  {journal} {\bibinfo  {journal} {Reports on Progress in Physics}\ }\textbf {\bibinfo {volume} {79}},\ \bibinfo {pages} {064501} (\bibinfo {year} {2016})}\BibitemShut {NoStop}%
\bibitem [{\citenamefont {Nakatsuji}\ \emph {et~al.}(2004)\citenamefont {Nakatsuji}, \citenamefont {Pines},\ and\ \citenamefont {Fisk}}]{NPF_model}%
  \BibitemOpen
  \bibfield  {author} {\bibinfo {author} {\bibfnamefont {S.}~\bibnamefont {Nakatsuji}}, \bibinfo {author} {\bibfnamefont {D.}~\bibnamefont {Pines}}, \ and\ \bibinfo {author} {\bibfnamefont {Z.}~\bibnamefont {Fisk}},\ }\href {\doibase 10.1103/PhysRevLett.92.016401} {\bibfield  {journal} {\bibinfo  {journal} {Phys. Rev. Lett.}\ }\textbf {\bibinfo {volume} {92}},\ \bibinfo {pages} {016401} (\bibinfo {year} {2004})}\BibitemShut {NoStop}%
\bibitem [{\citenamefont {Amorese}\ \emph {et~al.}(2020)\citenamefont {Amorese}, \citenamefont {Sundermann}, \citenamefont {Leedahl}, \citenamefont {Marino}, \citenamefont {Takegami}, \citenamefont {Gretarsson}, \citenamefont {Gloskovskii}, \citenamefont {Schlueter}, \citenamefont {Haverkort}, \citenamefont {Huang}, \citenamefont {Szlawska}, \citenamefont {Kaczorowski}, \citenamefont {Ran}, \citenamefont {Maple}, \citenamefont {Bauer}, \citenamefont {Leithe-Jasper}, \citenamefont {Hansmann}, \citenamefont {Thalmeier}, \citenamefont {Tjeng},\ and\ \citenamefont {Severing}}]{Amorese_2020}%
  \BibitemOpen
  \bibfield  {author} {\bibinfo {author} {\bibfnamefont {A.}~\bibnamefont {Amorese}}, \bibinfo {author} {\bibfnamefont {M.}~\bibnamefont {Sundermann}}, \bibinfo {author} {\bibfnamefont {B.}~\bibnamefont {Leedahl}}, \bibinfo {author} {\bibfnamefont {A.}~\bibnamefont {Marino}}, \bibinfo {author} {\bibfnamefont {D.}~\bibnamefont {Takegami}}, \bibinfo {author} {\bibfnamefont {H.}~\bibnamefont {Gretarsson}}, \bibinfo {author} {\bibfnamefont {A.}~\bibnamefont {Gloskovskii}}, \bibinfo {author} {\bibfnamefont {C.}~\bibnamefont {Schlueter}}, \bibinfo {author} {\bibfnamefont {M.~W.}\ \bibnamefont {Haverkort}}, \bibinfo {author} {\bibfnamefont {Y.}~\bibnamefont {Huang}}, \bibinfo {author} {\bibfnamefont {M.}~\bibnamefont {Szlawska}}, \bibinfo {author} {\bibfnamefont {D.}~\bibnamefont {Kaczorowski}}, \bibinfo {author} {\bibfnamefont {S.}~\bibnamefont {Ran}}, \bibinfo {author} {\bibfnamefont {M.~B.}\ \bibnamefont {Maple}}, \bibinfo {author} {\bibfnamefont {E.~D.}\ \bibnamefont {Bauer}}, \bibinfo {author} {\bibfnamefont
  {A.}~\bibnamefont {Leithe-Jasper}}, \bibinfo {author} {\bibfnamefont {P.}~\bibnamefont {Hansmann}}, \bibinfo {author} {\bibfnamefont {P.}~\bibnamefont {Thalmeier}}, \bibinfo {author} {\bibfnamefont {L.~H.}\ \bibnamefont {Tjeng}}, \ and\ \bibinfo {author} {\bibfnamefont {A.}~\bibnamefont {Severing}},\ }\href {\doibase 10.1073/pnas.2005701117} {\bibfield  {journal} {\bibinfo  {journal} {Proceedings of the National Academy of Sciences}\ }\textbf {\bibinfo {volume} {117}},\ \bibinfo {pages} {30220} (\bibinfo {year} {2020})}\BibitemShut {NoStop}%
\bibitem [{\citenamefont {Lee}\ \emph {et~al.}(2018)\citenamefont {Lee}, \citenamefont {Matsuda}, \citenamefont {Mydosh}, \citenamefont {Zaliznyak}, \citenamefont {Kolesnikov}, \citenamefont {S\"ullow}, \citenamefont {Ruff},\ and\ \citenamefont {Granroth}}]{Lee_2018}%
  \BibitemOpen
  \bibfield  {author} {\bibinfo {author} {\bibfnamefont {J.}~\bibnamefont {Lee}}, \bibinfo {author} {\bibfnamefont {M.}~\bibnamefont {Matsuda}}, \bibinfo {author} {\bibfnamefont {J.~A.}\ \bibnamefont {Mydosh}}, \bibinfo {author} {\bibfnamefont {I.}~\bibnamefont {Zaliznyak}}, \bibinfo {author} {\bibfnamefont {A.~I.}\ \bibnamefont {Kolesnikov}}, \bibinfo {author} {\bibfnamefont {S.}~\bibnamefont {S\"ullow}}, \bibinfo {author} {\bibfnamefont {J.~P.~C.}\ \bibnamefont {Ruff}}, \ and\ \bibinfo {author} {\bibfnamefont {G.~E.}\ \bibnamefont {Granroth}},\ }\href {\doibase 10.1103/PhysRevLett.121.057201} {\bibfield  {journal} {\bibinfo  {journal} {Phys. Rev. Lett.}\ }\textbf {\bibinfo {volume} {121}},\ \bibinfo {pages} {057201} (\bibinfo {year} {2018})}\BibitemShut {NoStop}%
\bibitem [{\citenamefont {Thomas}\ \emph {et~al.}(2020)\citenamefont {Thomas}, \citenamefont {Santos}, \citenamefont {Christensen}, \citenamefont {Asaba}, \citenamefont {Ronning}, \citenamefont {Thompson}, \citenamefont {Bauer}, \citenamefont {Fernandes}, \citenamefont {Fabbris},\ and\ \citenamefont {Rosa}}]{Thomas_2020}%
  \BibitemOpen
  \bibfield  {author} {\bibinfo {author} {\bibfnamefont {S.~M.}\ \bibnamefont {Thomas}}, \bibinfo {author} {\bibfnamefont {F.~B.}\ \bibnamefont {Santos}}, \bibinfo {author} {\bibfnamefont {M.~H.}\ \bibnamefont {Christensen}}, \bibinfo {author} {\bibfnamefont {T.}~\bibnamefont {Asaba}}, \bibinfo {author} {\bibfnamefont {F.}~\bibnamefont {Ronning}}, \bibinfo {author} {\bibfnamefont {J.~D.}\ \bibnamefont {Thompson}}, \bibinfo {author} {\bibfnamefont {E.~D.}\ \bibnamefont {Bauer}}, \bibinfo {author} {\bibfnamefont {R.~M.}\ \bibnamefont {Fernandes}}, \bibinfo {author} {\bibfnamefont {G.}~\bibnamefont {Fabbris}}, \ and\ \bibinfo {author} {\bibfnamefont {P.~F.~S.}\ \bibnamefont {Rosa}},\ }\href@noop {} {\bibfield  {journal} {\bibinfo  {journal} {Science Advances}\ }\textbf {\bibinfo {volume} {6}} (\bibinfo {year} {2020})}\BibitemShut {NoStop}%
\bibitem [{\citenamefont {Li}\ \emph {et~al.}(2021)\citenamefont {Li}, \citenamefont {Nakamura}, \citenamefont {Honda}, \citenamefont {Sato}, \citenamefont {Homma}, \citenamefont {Shimizu}, \citenamefont {Ishizuka}, \citenamefont {Yanase}, \citenamefont {Knebel}, \citenamefont {Flouquet},\ and\ \citenamefont {Aoki}}]{Akoi}%
  \BibitemOpen
  \bibfield  {author} {\bibinfo {author} {\bibfnamefont {D.}~\bibnamefont {Li}}, \bibinfo {author} {\bibfnamefont {A.}~\bibnamefont {Nakamura}}, \bibinfo {author} {\bibfnamefont {F.}~\bibnamefont {Honda}}, \bibinfo {author} {\bibfnamefont {Y.}~\bibnamefont {Sato}}, \bibinfo {author} {\bibfnamefont {Y.}~\bibnamefont {Homma}}, \bibinfo {author} {\bibfnamefont {Y.}~\bibnamefont {Shimizu}}, \bibinfo {author} {\bibfnamefont {J.}~\bibnamefont {Ishizuka}}, \bibinfo {author} {\bibfnamefont {Y.}~\bibnamefont {Yanase}}, \bibinfo {author} {\bibfnamefont {G.}~\bibnamefont {Knebel}}, \bibinfo {author} {\bibfnamefont {J.}~\bibnamefont {Flouquet}}, \ and\ \bibinfo {author} {\bibfnamefont {D.}~\bibnamefont {Aoki}},\ }\href@noop {} {\bibfield  {journal} {\bibinfo  {journal} {Journal of the Physical Society of Japan}\ }\textbf {\bibinfo {volume} {90}} (\bibinfo {year} {2021})}\BibitemShut {NoStop}%
\bibitem [{\citenamefont {Christovam}\ \emph {et~al.}(2024)\citenamefont {Christovam}, \citenamefont {Sundermann}, \citenamefont {Marino}, \citenamefont {Takegami}, \citenamefont {Falke}, \citenamefont {Dolmantas}, \citenamefont {Harder}, \citenamefont {amd Bernhard~Keimer}, \citenamefont {Gloskovskii}, \citenamefont {Haverkort}, \citenamefont {Elfimov}, \citenamefont {Zwicknagl}, \citenamefont {Andreev}, \citenamefont {Havela}, \citenamefont {Bordelon}, \citenamefont {Bauer}, \citenamefont {Rosa}, \citenamefont {Severing},\ and\ \citenamefont {Tjeng}}]{christovam2024stabilization}%
  \BibitemOpen
  \bibfield  {author} {\bibinfo {author} {\bibfnamefont {D.~S.}\ \bibnamefont {Christovam}}, \bibinfo {author} {\bibfnamefont {M.}~\bibnamefont {Sundermann}}, \bibinfo {author} {\bibfnamefont {A.}~\bibnamefont {Marino}}, \bibinfo {author} {\bibfnamefont {D.}~\bibnamefont {Takegami}}, \bibinfo {author} {\bibfnamefont {J.}~\bibnamefont {Falke}}, \bibinfo {author} {\bibfnamefont {P.}~\bibnamefont {Dolmantas}}, \bibinfo {author} {\bibfnamefont {M.}~\bibnamefont {Harder}}, \bibinfo {author} {\bibfnamefont {H.~G.}\ \bibnamefont {amd Bernhard~Keimer}}, \bibinfo {author} {\bibfnamefont {A.}~\bibnamefont {Gloskovskii}}, \bibinfo {author} {\bibfnamefont {M.~W.}\ \bibnamefont {Haverkort}}, \bibinfo {author} {\bibfnamefont {I.}~\bibnamefont {Elfimov}}, \bibinfo {author} {\bibfnamefont {G.}~\bibnamefont {Zwicknagl}}, \bibinfo {author} {\bibfnamefont {A.~V.}\ \bibnamefont {Andreev}}, \bibinfo {author} {\bibfnamefont {L.}~\bibnamefont {Havela}}, \bibinfo {author} {\bibfnamefont {M.~M.}\ \bibnamefont {Bordelon}}, \bibinfo
  {author} {\bibfnamefont {E.~D.}\ \bibnamefont {Bauer}}, \bibinfo {author} {\bibfnamefont {P.~F.~S.}\ \bibnamefont {Rosa}}, \bibinfo {author} {\bibfnamefont {A.}~\bibnamefont {Severing}}, \ and\ \bibinfo {author} {\bibfnamefont {L.~H.}\ \bibnamefont {Tjeng}},\ }\href {https://arxiv.org/abs/2402.03852} {\enquote {\bibinfo {title} {{Stabilization of U 5$f^2$ configuration in UTe$_2$ through U 6d dimers in the presence of Te2 chains}},}\ } (\bibinfo {year} {2024}),\ \Eprint {http://arxiv.org/abs/2402.03852} {arXiv:2402.03852 [cond-mat.str-el]} \BibitemShut {NoStop}%
\bibitem [{\citenamefont {Read}\ and\ \citenamefont {Newns}(1983)}]{NRead_1983}%
  \BibitemOpen
  \bibfield  {author} {\bibinfo {author} {\bibfnamefont {N.}~\bibnamefont {Read}}\ and\ \bibinfo {author} {\bibfnamefont {D.~M.}\ \bibnamefont {Newns}},\ }\href {\doibase 10.1088/0022-3719/16/17/014} {\bibfield  {journal} {\bibinfo  {journal} {Journal of Physics C: Solid State Physics}\ }\textbf {\bibinfo {volume} {16}},\ \bibinfo {pages} {3273} (\bibinfo {year} {1983})}\BibitemShut {NoStop}%
\bibitem [{\citenamefont {Scott}\ and\ \citenamefont {Kwasigroch}(2025)}]{Scott_2025}%
  \BibitemOpen
  \bibfield  {author} {\bibinfo {author} {\bibfnamefont {E.}~\bibnamefont {Scott}}\ and\ \bibinfo {author} {\bibfnamefont {M.}~\bibnamefont {Kwasigroch}},\ }\href {\doibase 10.1103/7zsl-4497} {\bibfield  {journal} {\bibinfo  {journal} {Phys. Rev. Res.}\ }\textbf {\bibinfo {volume} {7}},\ \bibinfo {pages} {043225} (\bibinfo {year} {2025})}\BibitemShut {NoStop}%
\bibitem [{\citenamefont {Peters}\ \emph {et~al.}(2012)\citenamefont {Peters}, \citenamefont {Kawakami},\ and\ \citenamefont {Pruschke}}]{Peters_2012}%
  \BibitemOpen
  \bibfield  {author} {\bibinfo {author} {\bibfnamefont {R.}~\bibnamefont {Peters}}, \bibinfo {author} {\bibfnamefont {N.}~\bibnamefont {Kawakami}}, \ and\ \bibinfo {author} {\bibfnamefont {T.}~\bibnamefont {Pruschke}},\ }\href {\doibase 10.1103/PhysRevLett.108.086402} {\bibfield  {journal} {\bibinfo  {journal} {Phys. Rev. Lett.}\ }\textbf {\bibinfo {volume} {108}},\ \bibinfo {pages} {086402} (\bibinfo {year} {2012})}\BibitemShut {NoStop}%
\bibitem [{\citenamefont {Wu}\ \emph {et~al.}(2025{\natexlab{d}})\citenamefont {Wu}, \citenamefont {Weinberger}, \citenamefont {Hickey}, \citenamefont {Chichinadze}, \citenamefont {Shaffer}, \citenamefont {Cabala}, \citenamefont {Chen}, \citenamefont {Long}, \citenamefont {Brumm}, \citenamefont {Xie}, \citenamefont {Ling}, \citenamefont {Zhu}, \citenamefont {Skourski}, \citenamefont {Graf}, \citenamefont {Sechovsk\'y}, \citenamefont {Vali\ifmmode~\check{s}\else \v{s}\fi{}ka}, \citenamefont {Lonzarich}, \citenamefont {Grosche},\ and\ \citenamefont {Eaton}}]{qcldata}%
  \BibitemOpen
  \bibfield  {author} {\bibinfo {author} {\bibfnamefont {Z.}~\bibnamefont {Wu}}, \bibinfo {author} {\bibfnamefont {T.~I.}\ \bibnamefont {Weinberger}}, \bibinfo {author} {\bibfnamefont {A.~J.}\ \bibnamefont {Hickey}}, \bibinfo {author} {\bibfnamefont {D.~V.}\ \bibnamefont {Chichinadze}}, \bibinfo {author} {\bibfnamefont {D.}~\bibnamefont {Shaffer}}, \bibinfo {author} {\bibfnamefont {A.}~\bibnamefont {Cabala}}, \bibinfo {author} {\bibfnamefont {H.}~\bibnamefont {Chen}}, \bibinfo {author} {\bibfnamefont {M.}~\bibnamefont {Long}}, \bibinfo {author} {\bibfnamefont {T.~J.}\ \bibnamefont {Brumm}}, \bibinfo {author} {\bibfnamefont {W.}~\bibnamefont {Xie}}, \bibinfo {author} {\bibfnamefont {Y.}~\bibnamefont {Ling}}, \bibinfo {author} {\bibfnamefont {Z.}~\bibnamefont {Zhu}}, \bibinfo {author} {\bibfnamefont {Y.}~\bibnamefont {Skourski}}, \bibinfo {author} {\bibfnamefont {D.~E.}\ \bibnamefont {Graf}}, \bibinfo {author} {\bibfnamefont {V.}~\bibnamefont {Sechovsk\'y}}, \bibinfo {author} {\bibfnamefont {M.}~\bibnamefont
  {Vali\ifmmode~\check{s}\else \v{s}\fi{}ka}}, \bibinfo {author} {\bibfnamefont {G.~G.}\ \bibnamefont {Lonzarich}}, \bibinfo {author} {\bibfnamefont {F.~M.}\ \bibnamefont {Grosche}}, \ and\ \bibinfo {author} {\bibfnamefont {A.~G.}\ \bibnamefont {Eaton}},\ }\href {http://doi.org/10.17863/CAM.116724} {\enquote {\bibinfo {title} {{Research data supporting: A quantum critical line bounds the high field metamagnetic transition surface in UTe$_2$}},}\ } (\bibinfo {year} {2025}{\natexlab{d}}),\ \bibinfo {note} {{University of Cambridge Apollo Repository}}\BibitemShut {NoStop}%
\end{thebibliography}%
\clearpage

\begin{figure*}[!htbp]
    \centering
    \includegraphics[width=1\linewidth]{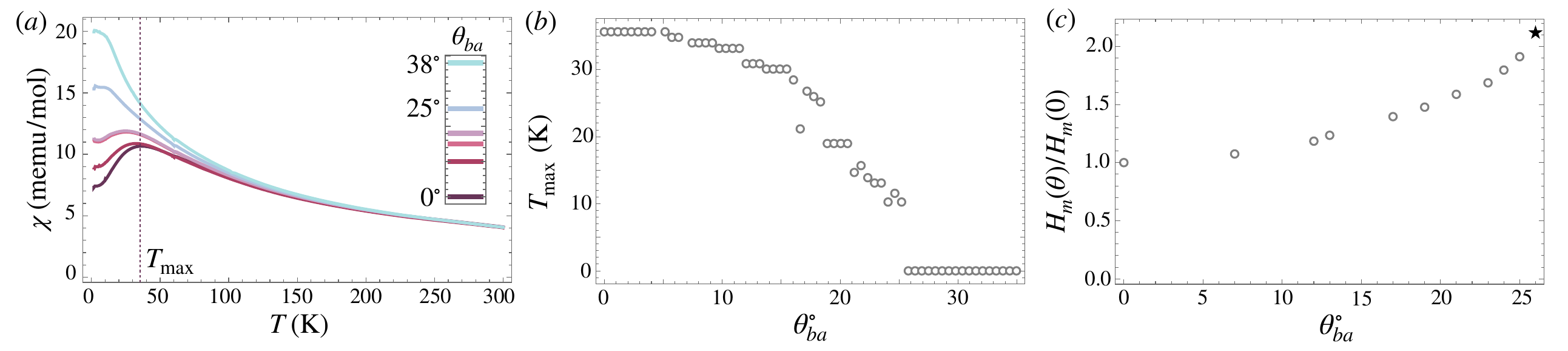}
    \caption{(a) The experimentally measured susceptibility as a function of temperature for field applied at an angle $\theta_{ba}$ from the $b$-axis towards the $a$-axis. (b) The experimentally measured temperature of the susceptibility maximum $T_{\rm max}$ as a function of $\theta_{ba}$. (c) The value of the critical field $h_m$ as a function of $\theta_{ba}$. The star denotes the critical angle where the metamagnetic jump vanishes. It is worth noting that the angle at which the metamagnetic jump vanishes is very close to the angle at which the maximum in the susceptibility disappears.}
    \label{fig:BA}
\end{figure*}
\appendix
\section{Results for magnetic field applied in the $ab$-plane}\label{app:ba}
Fig. \ref{fig:BA} shows the experimental results for the susceptibility, the susceptibility maximum temperature $T_{\rm max}$, and the critical field $h_m$ at which MMT takes places for magnetic field applied at an angle $\theta_{ba}$ from the $b$-axis towards the $a$-axis.
\end{document}